\documentclass[journal]{IEEEtran}
\usepackage{amsmath,amssymb,amsfonts, mathrsfs}
\usepackage{algorithmic}
\usepackage{algorithm}
\usepackage{array}
\ifCLASSOPTIONcompsoc
    \usepackage[caption=false, font=normalsize, labelfont=sf, textfont=sf]{subfig}
\else
\usepackage[caption=false, font=footnotesize]{subfig}
\fi
\usepackage{textcomp}

\usepackage{url}
\usepackage{graphicx}
\usepackage{epstopdf}
\usepackage{cite}
\usepackage{booktabs}
\usepackage{multirow}
\usepackage{bigstrut}
\usepackage{color}

\graphicspath{{images/}}
\usepackage{subfiles}

\begin{document}

\title{Real-time Cooperative Vehicle Coordination at Unsignalized Road Intersections}




\author{Jiping~Luo,
        Tingting~Zhang,~\IEEEmembership{Member,~IEEE,}
        Rui~Hao,
        Donglin~Li,
        Chunsheng~Chen,\\
        Zhenyu~Na,~\IEEEmembership{Member,~IEEE,}
        and~Qinyu~Zhang,~\IEEEmembership{Senior~Member,~IEEE}

}
\maketitle

\begin{abstract}
Cooperative coordination at unsignalized road intersections, which aims to improve the driving safety and traffic throughput for connected and automated vehicles (CAVs), has attracted increasing interests in recent years. However, most existing investigations either suffer from computational complexity or cannot harness the full potential of the road infrastructure. To this end, we first present a dedicated intersection coordination framework, where the involved vehicles hand over their control authorities and follow instructions from a centralized coordinator. Then a unified cooperative trajectory planning problem will be formulated to maximize the traffic throughput while ensuring driving safety. To address the key computational challenges in the real-world deployment, we reformulate this non-convex sequential decision-making problem into a model-free Markov Decision Process (MDP) and tackle it by devising a Twin Delayed Deep Deterministic Policy Gradient (TD3)-based strategy in the deep reinforcement learning (DRL) framework. Simulation and practical experiments show that the proposed strategy could achieve near-optimal performance in sub-static coordination scenarios and significantly improve the traffic throughput in the realistic continuous traffic flow. The most remarkable advantage is that our strategy could reduce the time complexity of computation to milliseconds, and is shown scalable when the road lanes increase.
\end{abstract}

\section{INTRODUCTION}
\IEEEPARstart{W}{ith} the proliferation of traffic demand and vehicle ownership, the current transportation systems are now facing some serious social problems: traffic congestion, accidents, fuel consumption, air pollution, etc\cite{traffic_1, traffic_2, traffic_3, traffic_4, traffic_5}. Road intersections where multiple roads cross and merge are the main bottlenecks for urban traffic. As reported in \cite{congesition_1, accident}, congestion and accidents in these areas cause substantial economic loss and tremendous safety hazards to society, and excessive travel time for drivers. Therefore, it is critical to develop dedicated intersection management systems, to allow vehicles to cross the intersection safely and efficiently.

Traditional traffic signal control (TSC) systems, where the traffic signal timing is predetermined according to historical traffic data, is widely used in current intersection coordination systems. However, this solution may not accommodate real-time traffic flow and bursty traffic demands \cite{TSC_2}. Benefiting from the artificial intelligence, especially reinforcement learning (RL), the adaptive TSC which could adjust traffic signal timing according to real-time traffic demand, has been shown effectiveness in reducing traffic congestion \cite{TSC_0,TSC_1,TSC_2,TSC_3,TSC_4,DRL_ITS}. However, this paradigm is still far from satisfactory with respect to both driving safety and traffic efficiency as over 90 percent of crashes are tied to improper human behaviors \cite{accident}. Moreover, the inconveniences of frequent stops and idling at intersections not only waste both fuel and time but also degrade the comfort of drivers and passengers \cite{comfort, coordination_CAV_survey,chenlei_1}.

In recent years, centralized coordination at unsignalized intersections has been widely studied to improve the performance of intersection control, especially when interoperations are required among vehicles. In this paradigm, there usually exists a coordination node that is fulfilled at the road side unit (RSU) to collect state information such as position, velocity, acceleration, heading angle, and intention of all involved vehicles via vehicle-to-infrastructure (V2I) communication links. After proper coordination algorithms, the maneuver instructions could be generated and forwarded to vehicles. Recent investigations show that centralized coordinators are advantageous in traffic throughput, collision-free, deadlock-free, passenger comfort, fuel efficiency, etc \cite{comfort,coordination_CAV_survey,chenlei_1,caodongpu_survey, jiapan}, making them suitable for real-world deployments. However, the design of centralized coordination algorithms confronts challenges due to the collision-free constraints and the complexity of the required maneuvers.

Lee and Park\cite{safety_gap} proposed a cooperative vehicle control strategy for connected and automated vehicles to reduce stop delay, total passing time, and fuel consumption. They adopted the following two strict assumptions: (1) Any two vehicles with overlapped paths should not exist at the intersection area at the same time, thereby eliminating potential collisions coming from all conflicting approaches. (2) The acceleration of each vehicle is fixed from its initial position to the end of the intersection. Though this algorithm could improve coordination performance compared with traditional TSC, it is apparent that the above two assumptions would impose limitations on the maximum coordination capacities of the intersection. Moreover, the formulated nonlinear constrained programming (NCP) problem was solved by employing search-based algorithms such as the Active Set Method (ASM), Interior Point Method (IPM) and genetic algorithm (GA) that could produce undesirable solutions and computation time.

The authors in\cite{efficient_algorithm, hult_1, hult_2, moyangan, CS, liuchanghao} adopted similar rules that the whole intersection area was reserved for conflicting vehicles one after another, namely \textit{collision set} (or \textit{critical set}) strategy. The most notable difference is that the acceleration is treated as a continuous variable rather than a fixed value for the subsequent time slots, which could further improve the coordination efficiency and passenger comfort. However, due to the non-convexity of the formulated problems, some relaxations such as the mixed integer programming (MIP) solutions are required, which inevitably increase computational complexity. In\cite{CS, liuchanghao}, the stability of the coordination system was also considered as a metric to trade off long-term against short-term performances. Aiming at the computational complexity, appropriate task offloading from the centralized coordinator was shown to be an option \cite{moyanganwcncw}.

To address the underutilization issue of the intersection spatial resources, Kamal et al.\cite{CCP} defined the cross-collision points (CCPs) inside the intersection area where the paths of any two conflicting vehicles intersect. The circle (with a certain radius) centering the CCP shows an approximate area where two vehicles are not allowed to enter at the same time. In addition, smooth traffic flows were achieved by solving an NCP problem at each time step over a finite horizon in the model predictive control (MPC) framework. However, as stated in this paper, MPC is usually computationally demanding and the average computation time is about 1.76 seconds per iteration, which is impractical in real-world deployments. Similarly, the intersection area can also be divided into multiple rectangular sub-zones (each could be reserved for one vehicle at a time) according to the combination of each pair of orthogonal lanes\cite{comfort, lili, lili_2}. However, this reservation rule depends heavily on the road structure and requires much more safety redundancy than vehicle geometry.

More recently, the authors in\cite{topological_braids} and\cite{STRS} brought totally different insights that the cooperative trajectories can be optimized in the three-dimensional space with X, Y and T domains to harness the full potential of the intersection. In\cite{STRS}, a space-time resource searching (STRS) algorithm was proposed to search for compact and disjoint trajectories in the XYT domains, which allows multiple vehicles to share the intersection simultaneously and safely without any reservation rules, thereby providing tremendous coordination efficiency advantages over the aforementioned rule-based strategies. However, the STRS strategy shows obvious disadvantages in real-time processing performance due to the high dimensional searching space and the combination of dynamic programming (DP) and quadratic programming (QP).

In this paper, comprehensively considering the key issues, i.e., traffic throughput, driving safety, and computational complexity, at unsignalized intersections, a Twin Delayed Deep Deterministic Policy Gradient (TD3)-based centralized coordination strategy is designed in the deep reinforcement learning (DRL) framework. The contributions of this paper can be summarized as follows.
\begin{itemize}
  \item Leveraging \textit{Lyapunov Optimization Theorem}, a unified cooperative trajectory planning problem is formulated to maximize traffic throughput while ensuring driving safety. To harness the full coordination potential of the road intersections, we adopt the paradigm that all controlled vehicles could occupy the intersection simultaneously without any reservation rules.
  \item To rapidly solve the formulated non-convex sequential decision problem and obtain optimal disjoint trajectories in XYT domains, we transform this problem into a model-free Markov Decision Process (MDP) and elucidate the design of the state space, action space and reward function in detail. Then a TD3-based centralized coordinator is trained offline and can be deployed online to meet the stringent real-time coordination requirements.
  \item Simulation results and laboratorial experiments demonstrate that our TD3-based strategy could achieve near-optimal performance in static coordination scenarios and significantly improve traffic throughput in continuous traffic flow. The most impressive advantage is that our strategy has millisecond computational latency, which outperforms most existing centralized strategies by which the computational complexity increases exponentially with the number of vehicles and lanes. Moreover, our strategy shows good adaptability in low and medium traffic loads compared with traditional TSC systems.
\end{itemize}

The rest of this paper is organized as follows. First, the system model is presented in Section \ref{system_model}, including the centralized coordination framework, formulation of optimal coordination problem, and benchmark solutions. In Section \ref{solution}, the formulated optimization problem is transformed and solved by devising a TD3-based strategy. Section \ref{simulation_results} presents extensive simulation results and experiments to confirm the capability and applicability of the proposed approach. Finally, the conclusion and future work are presented in Section \ref{conclusion}.

\section{System Model}\label{system_model}

\subsection{Intersection Coordination Framework}
\begin{figure}[h]
    \centering
    \includegraphics[scale=0.32]{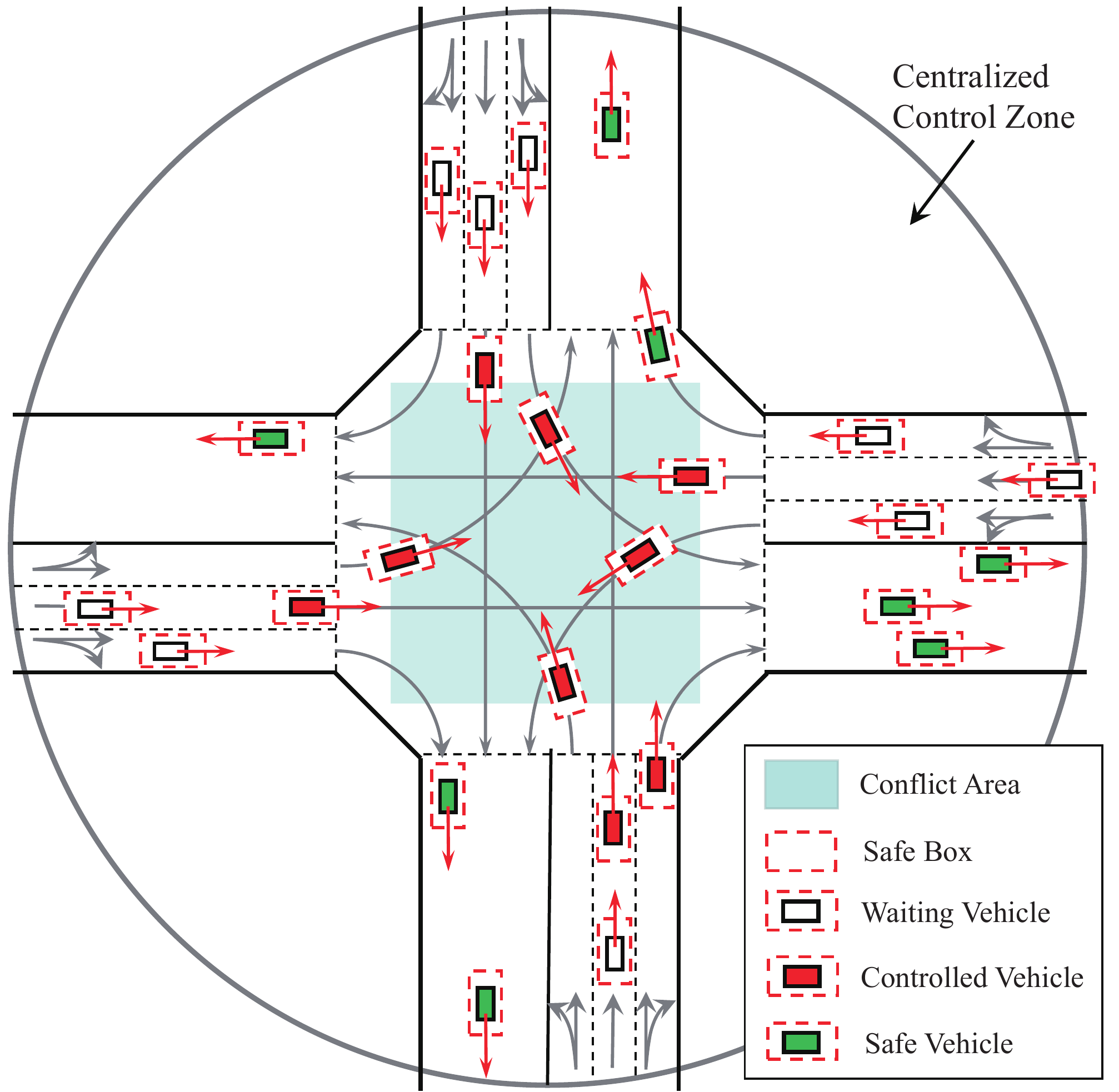}
    \caption{Graphic model of the intersection coordination system.}
    \label{fig:intersection}
\end{figure}

We consider a typical unsignalized intersection which consists of $R$ roads (usually we set $R = 4$), each with $K$ lanes, as shown in Fig. \ref{fig:intersection}. The area within the circle is called the Centralized Control Zone (CCZ) where a centralized coordinator is deployed at the road side unit (RSU) to collect the state information of the vehicles and give instructions.
The blue area is called the Conflict Area (CA) where a potential collision may occur.

In the CCZ, all vehicles will go through three different phases, i.e., waiting vehicle (WV), controlled vehicle (CV), and safe vehicle (SV).
Once a vehicle enters the CCZ, it will travel at a low speed $v_\text{low}$ and wait for coordination, namely WV. The WVs need to keep a safe distance from the vehicles ahead and cannot enter the CA without permission.
CVs represent the vehicles that update their state information to the centralized coordinator and need to leave the CCZ under instructions.
SVs are the CVs that have passed the CA without collision. If all the CVs turn green, the coordinator can focus on the next batch of CVs, and all SVs can leave the CCZ at a specified speed along their respective paths.

In the decision-making issue of unsignalized intersections, the large number of vehicles, high collision probability, and the inherent need for real-time processing make the centralized coordination system difficult to implement. For ease of analysis, we adopt the following assumptions:
\begin{itemize}
    \item [1)] The vehicles in the leftmost/rightmost lane of each road can turn left/right or go straight, and the rest can only go straight. Changing lane maneuvers is prohibited in the CCZ to ensure driving safety.
    \item [2)] Each vehicle shares its driving states (position, velocity, etc.) and intentions (i.e., go straight, turn left and turn right) with the centralized coordinator via vehicle-to-infrastructure (V2I) communication links. Based on the collected information, the coordinator could broadcast the computed instructions (acceleration, steering angle, etc.) to the CVs without any error or packet drops.
    \item [3)] Without loss of generality, each vehicle travels along its reference path which is predefined and fixed.
    \item [5)] The coordinator adopts a batch optimization approach, where each batch contains up to $NRK$ CVs (i.e., each lane has up to $N$ CVs). The rationality of this approach will be demonstrated in Section \ref{sec:problem_formulation}.
\end{itemize}

Notations $\mathcal{R} = \{1, 2, \ldots, R\}$, $\mathcal{K}_r = \{1, 2, \ldots, K\}$ and $\mathcal{N}_{r,k} = \{1, 2, \ldots, N\}$ represent the set of roads, the set of lanes of road $r$ and the set of CVs in the $k$-th lane of road $r$, respectively. Denote $\mathcal{I}_{r, k}/I_{r, k}$ as the set/number of vehicles in the $k$-th lane of road $r$. We use tuple $(r, k)$ and $(r, k, n)$ to indicate the $k$-th lane of road $r$ and the $n$-th CV in the $k$-th lane of road $r$, respectively. $(r_1, k_1, n_1) \neq (r_2, k_2, n_2)$ means two different CVs.

\subsection{Vehicle Kinematic Model}
\begin{figure}[htb]
    \centering
    \includegraphics[scale=0.35]{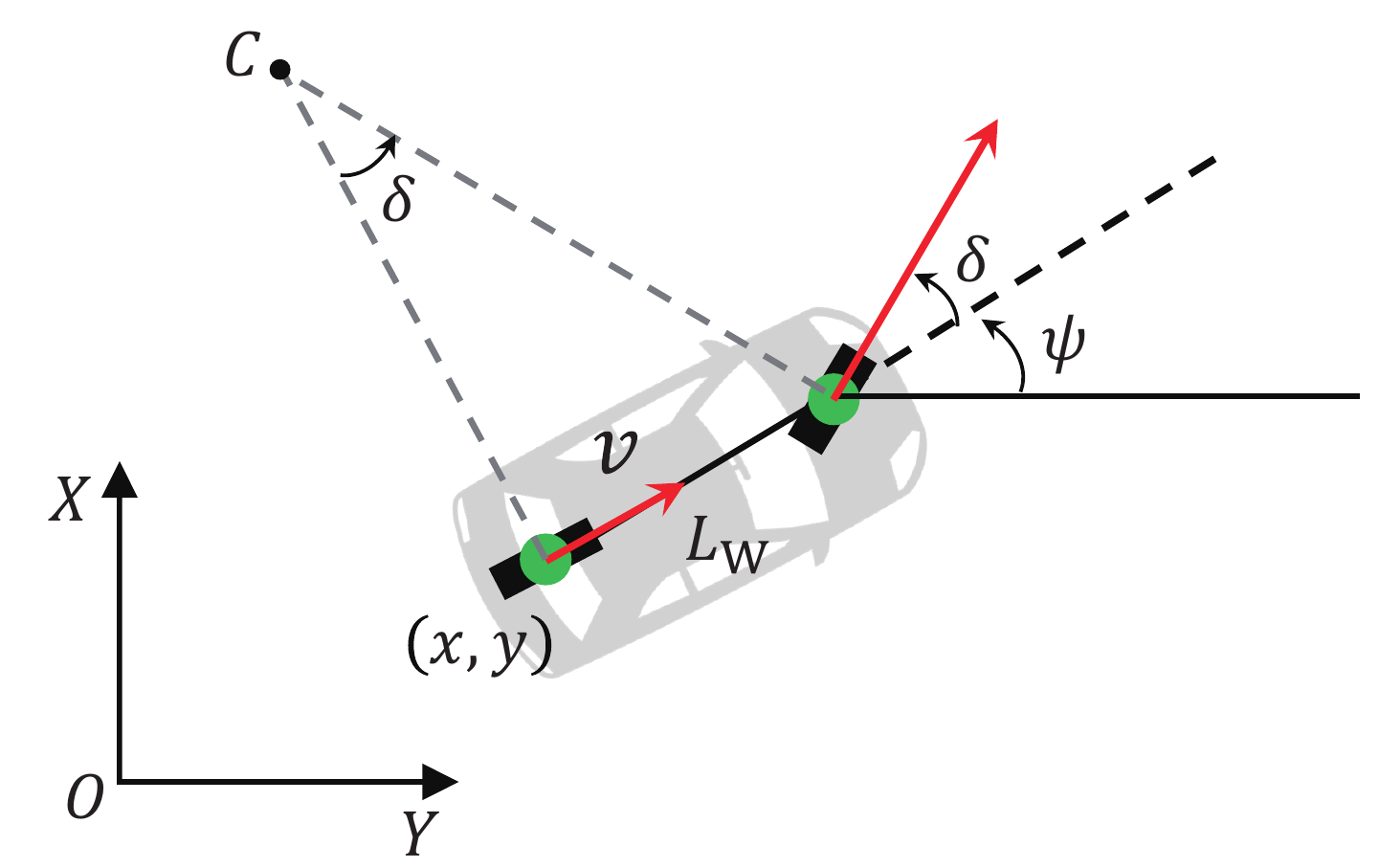}
    \caption{Schematics on the kinematic model of a front steering vehicle.}
    \label{fig:VKM}
\end{figure}
Since vehicles usually pass the intersection at middle or even low speeds and in order to reduce the computational complexity of the coordination algorithm, the popular kinematic bicycle model\cite{kinimatic} is adopted to control the vehicles in this paper, as shown in Fig. \ref{fig:VKM}. The state vector of the CV $(r, k, n)$ can be expressed as $\mathbf{z}_{r, k}^n = [x_{r, k}^n, y_{r, k}^n, v_{r, k}^n, \psi_{r, k}^n]^\text{T}$, where $(x_{r, k}^n, y_{r, k}^n)$, $v_{r, k}^n$ and $\psi_{r,k}^n$ are the position, longitudinal velocity and heading angle, respectively. The control input vector is defined as $[a_{r, k}^n, \delta_{r, k}^n]^\text{T}$, where $a_{r, k}^n$ is the longitudinal acceleration and $\delta_{r, k}^n$ is the steering angle. The vehicle dynamics are given as follows:
\begin{align}
    \begin{bmatrix}
        \dot{x}_{r, k}^n(t)      \\
        \dot{y}_{r, k}^n(t)    \\
        \dot{v}_{r, k}^n(t)    \\
        \dot{\psi}_{r, k}^n(t)  \\
    \end{bmatrix} =
    \begin{bmatrix}
        v_{r, k}^n(t)  \cos(\psi_{r, k}^n(t) )       \\
        v_{r, k}^n(t)  \sin(\psi_{r, k}^n(t) )       \\
        a_{r, k}^n(t)                           \\
        v_{r, k}^n(t)  \tan(\delta_{r, k}^n(t) ) / L_\text{w} \\
    \end{bmatrix} \label{eq:dynamic}
\end{align}
where $r \in \mathcal{R}, k \in \mathcal{K}_r, n \in \mathcal{N}_{r, k}$, $\dot{(\bullet)} := \frac{\partial}{\partial t}(\bullet)$, $L_\text{w}$ is the distance between the front and rear wheel axles. Considering the practical constraints in the vehicle model, we do have
\begin{align}
    - a_\text{max} & \leq  a_{r,k}^n(t)   \leq a_\text{max} \label{a}\\
       0     & \leq  v_{r, k}^n(t)   \leq v_\text{max} \label{v}\\
    -\delta_\text{max} & \leq \delta_{r, k}^n(t)  \leq \delta_\text{max} \label{delta}
\end{align}

In the three-dimensional Cartesian coordinates XYT, the trajectory function can be described as:
\begin{align}
    \begin{cases}
        \text{Path:}
            & \begin{cases}
                x_{r, k}^n = f_{r, k}^n(s_{r,k}^n) \\
                y_{r, k}^n = g_{r, k}^n(s_{r,k}^n)
              \end{cases} \\
        \text{Speed Profile:}  & s_{r,k}^n =  u_{r, k}^n(t)
    \end{cases} \label{eq:trajectory}
\end{align}
where $s$ is the arc length along the path, and $u$ is a function of velocity and acceleration. Given a reference path function $(f, g)$, e.g., line \& circle curves, polynomial curves, B\'{e}zier curves, and splines, the trajectory could be determined by optimizing the speed profile.


\subsection{Collision Detection}
\begin{figure}[htb]
    \centering
    \includegraphics[scale=0.28]{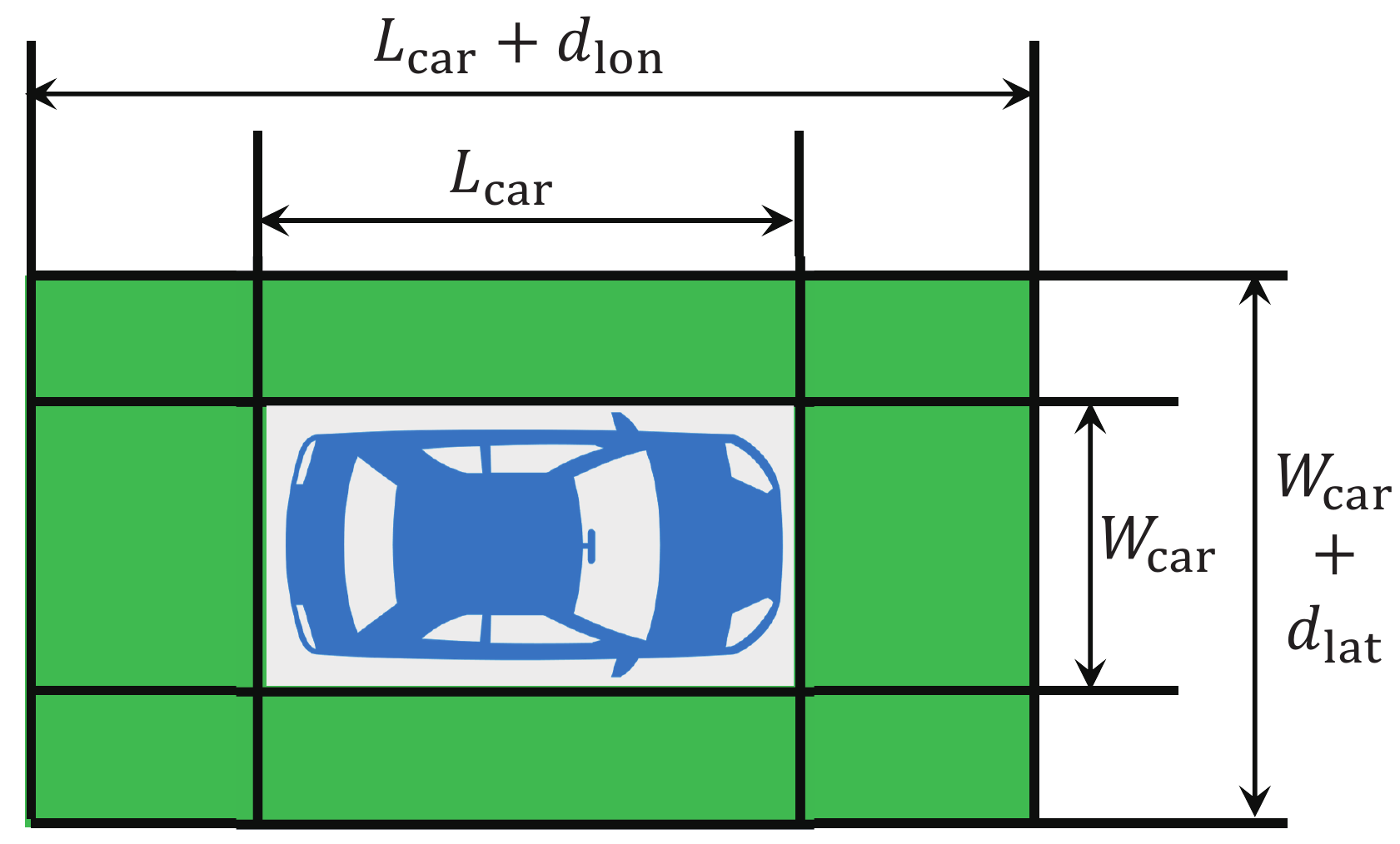}
    \caption{The occupied road region is modeled as a rectangle with safety redundancy.}
    \label{fig:occupy}
\end{figure}

In practice, there exist some imperfections in the coordination system. First, the noise in position and control could result in the vehicle deviating from the planned trajectory. Second, the communication delay is inevitable, which may reduce the freshness of the coordination instructions. To this end, we provide \textit{safety redundancy} for each vehicle to ensure the safety and robustness of the coordination system, as depicted in Fig. \ref{fig:occupy}.
The road region occupied by the vehicle is defined by a two-dimensional convex polygon $\mathcal{P}$, $L_\text{car}$ and $W_\text{car}$ are the length and width of the vehicle, $d_\text{lon}$ and $d_\text{lat}$ are the longitudinal and lateral safe distances.

The convex set $\mathcal{P}$ for a state vector $\mathbf{z} = [x, y, v, \psi]$ can be mathematically defined as\cite{occupied_road_region}:
\begin{align}
    \mathcal{P}(\mathbf{z}) &= \left\{
    \mathbf{p} \in \mathbb{R}^2 \big| \mathbf{A}(\mathbf{z}) \mathbf{p} \leq \mathbf{b}(\mathbf{z}) \right\} \\
    \mathbf{A}(\mathbf{z}) &=
    [
          \mathbf{M}(\psi)^\text{T},
         - \mathbf{M}(\psi)^\text{T}
    ]^\text{T} \\
    \mathbf{b}(\mathbf{z}) &= [L/2, W/2, L/2, W/2]^\text{T} + \mathbf{A}(\mathbf{z}) [x_\text{c}, y_\text{c}]^\text{T}
\end{align}
where $\mathbf{M}(\psi) = \begin{bmatrix}
    \cos(\psi) & -\sin (\psi) \\
    \sin (\psi) & \cos (\psi)
\end{bmatrix}$ is the rotation matrix, $L = L_\text{car} + d_\text{lon}$ and $W = W_\text{car} + d_\text{lat}$ are the length and width of the occupied road region, $x_\text{c} = x + L_\text{w}\cos(\psi)/2$ and $y_\text{c} = y + L_\text{w}\sin(\psi)/2$ are the coordinates of the center of the occupied road region. Collisions among the CVs can be avoided if the following constraints are satisfied throughout the coordination,
\begin{align}
    \mathcal{P}(z_{r_1, k_1}^{n_1}) \cap \mathcal{P}(z_{r_2, k_2}^{n_2}) = \varnothing, \forall (r_1, k_1, n_1) \neq (r_2, k_2, n_2)
    \label{eq:collision}
\end{align}
In this paper, collision detection is performed by leveraging Separating Axis Theorem (SAT)\cite{SAT}, which is known as an efficient algorithm for checking the overlapping between polygons. 

The occupied road region method is conservative as the size is pre-determined according to system noise magnitudes. Some advanced control algorithms, such as control barrier function (CBF)\cite{CBF} and covariance steering\cite{covariance_steering}, could readily adjust safety redundancy to compensate for possible measurement noise and modeling inaccuracy. In this paper, we focus on coordination efficiency and real-time processing capability, how to incorporate flexible safety redundancy in intersection coordination will be further explored in our future works.

\subsection{Problem Formulation}\label{sec:problem_formulation}
For intersection coordination system, the queue length of each lane, i.e., $I_{r, k}(t)$, is time-varying due to the dynamic arrival of vehicles.
Future queue states can be driven by stochastic arrival $a_{r, k}(t)$ and coordination rate $b_{r, k}(t)$ according to the following dynamic equation:
\begin{align}
    I_{r, k}(t+1) = \max \left[ I_{r, k}(t) + a_{r, k}(t) - b_{r, k}(t), 0 \right] \label{eq:queue}
\end{align}
The real-time coordination rate is defined as:
\begin{align}
     b_{r, k}(t) = \frac{N_{r, k}(t)}{T_\text{batch}} \label{rate}
\end{align}
where $0 \leq N_{r, k}(t) \leq N$ is the number of CVs in lane $(r, k)$ of current batch, $T_\text{batch} = \max \{T_{r, k} | r \in \mathcal{R}, k \in \mathcal{K}_r \}$ is the coordination time (the time elapsed from turning red to turning green) of current batch. Here, $T_{r, k} = \max \{T_{r, k}^n (\mathbf{u}_{r, k}^n) | n \in \mathcal{N}_{r,k}\}$ is the coordination time of lane $(r, k)$, and $\mathbf{u}_{r, k}^n$ is the speed profile of CV $(r, k, n)$.

Let $\mathbf{I}(t) = [I_{1, 1}(t), I_{1, 2}(t), \ldots, I_{R, K}(t)]^\text{T}$ be the vector of current queue backlogs, the \textit{Lyapunov function} \cite{lyapunov} can be defined as:
\begin{align}
    L(\mathbf{I}(t)) \triangleq \frac{1}{RK} \sum_{r \in \mathcal{R}}
    \sum_{k \in \mathcal{K}_r} I_{r, k}^2 (t) \label{eq:Lfunction}
\end{align}
To consistently push the Lyapunov function towards a low congestion region, we can minimize the bound on $\Delta (\mathbf{I}(t))$. Herein, $\Delta (\mathbf{I}(t))$ is the \textit{conditional Lyapunov drift} and can be defined as:
\begin{align}
    \Delta (\mathbf{I}(t)) &\triangleq \mathbb{E}\left\{ L(\mathbf{I}(t + 1)) - L(\mathbf{I}(t)) \big| \mathbf{I}(t)  \right\} \label{eq:drift}
\end{align}
Substitute (\ref{eq:queue}) and (\ref{eq:Lfunction}) into (\ref{eq:drift}), the bound can be derived as:
\begin{align}
    \Delta (\mathbf{I}(t))
    & \leq \frac{1}{{RK}}\mathbb{E}\left\{\sum_{r \in \mathcal{R}} \sum_{k \in \mathcal{K}_r} a_{r, k}(t)^2 + b_{r, k}(t)^2 \big| \mathbf{I}(t)  \right\} \notag\\
    & \, \, \, \, + \frac{2}{RK} \sum_{r \in \mathcal{R}} \sum_{k \in \mathcal{K}_r}I_{r, k}(t)\lambda_{r, k}  \notag \\
    & \, \, \, \, -
    \frac{2}{RK}\mathbb{E}\left\{\sum_{r \in \mathcal{R}} \sum_{k \in \mathcal{K}_r}I_{r, k}(t)b_{r, k}(t) \big| \mathbf{I}(t)
    \right\}
\end{align}
where $\lambda_{r, k} = \mathbb{E}\left\{ a_{r, k}(t) |\mathbf{I}(t) \right\}$ is the mean arrival rate. The first term on the right-hand-side of the above drift inequality can be upper bounded by a finite constant.
Therefore, minimize the bound on the drift is equivalent to maximize the following expression:
\begin{align}
    \mathbb{E}\left\{\sum_{r \in \mathcal{R}} \sum_{k \in \mathcal{K}_r}I_{r, k}(t)b_{r, k}(t) \big| \mathbf{I}(t) \right\} \label{eq:bound}
\end{align}

Expression (\ref{eq:bound}) provides a general mathematical form for the goal in the field of centralized intersection coordination. By using the concept of \textit{opportunistically maximizing an expectation}, (\ref{eq:bound}) can be maximized by maximizing the weighted sum of the coordination rate, i.e., $\sum_{r \in \mathcal{R}} \sum_{k \in \mathcal{K}_r}I_{r, k}(t)b_{r, k}(t)$. It follows that the CVs in the lane with more vehicles have higher priority and need to be provided with more coordination rate.
How to efficiently coordinate the CVs according to queue states to support higher coordination rate is important to the intersection coordination system. To achieve this goal, the optimization problem can be formulated as follows,
\begin{align*}
    \mathscr{P} : \, \,  \max_{\mathbf{u}_{r, k}^n} \, \, & \sum_{r \in \mathcal{R}} \sum_{k \in \mathcal{K}_r}I_{r, k}(t)b_{r, k}(t) \\
    \mathrm{\mathbf{s.t.}} \, \,
         & \, (\ref{eq:dynamic})-(\ref{eq:trajectory}), \, (\ref{eq:collision}) - (\ref{rate})
\end{align*}
For a given initial configuration, a multitude of solutions might exist (i.e., when and in what control sequence each CV cross the intersection), and the optimal trajectories can only be found by a structured exploration of the different alternatives\cite{hult_2}. Thus, the problem $\mathscr{P}$ is NP-hard in general.

\subsection{Benchmark Solutions}\label{sec:benchmarks}
In this subsection, we briefly introduce two state-of-the-art centralized intersection coordination solutions as benchmarks.

\subsubsection{Collision Set}
\begin{figure}[t]
    \centering
    \subfloat[collision set]{%
        \includegraphics[scale=0.2]{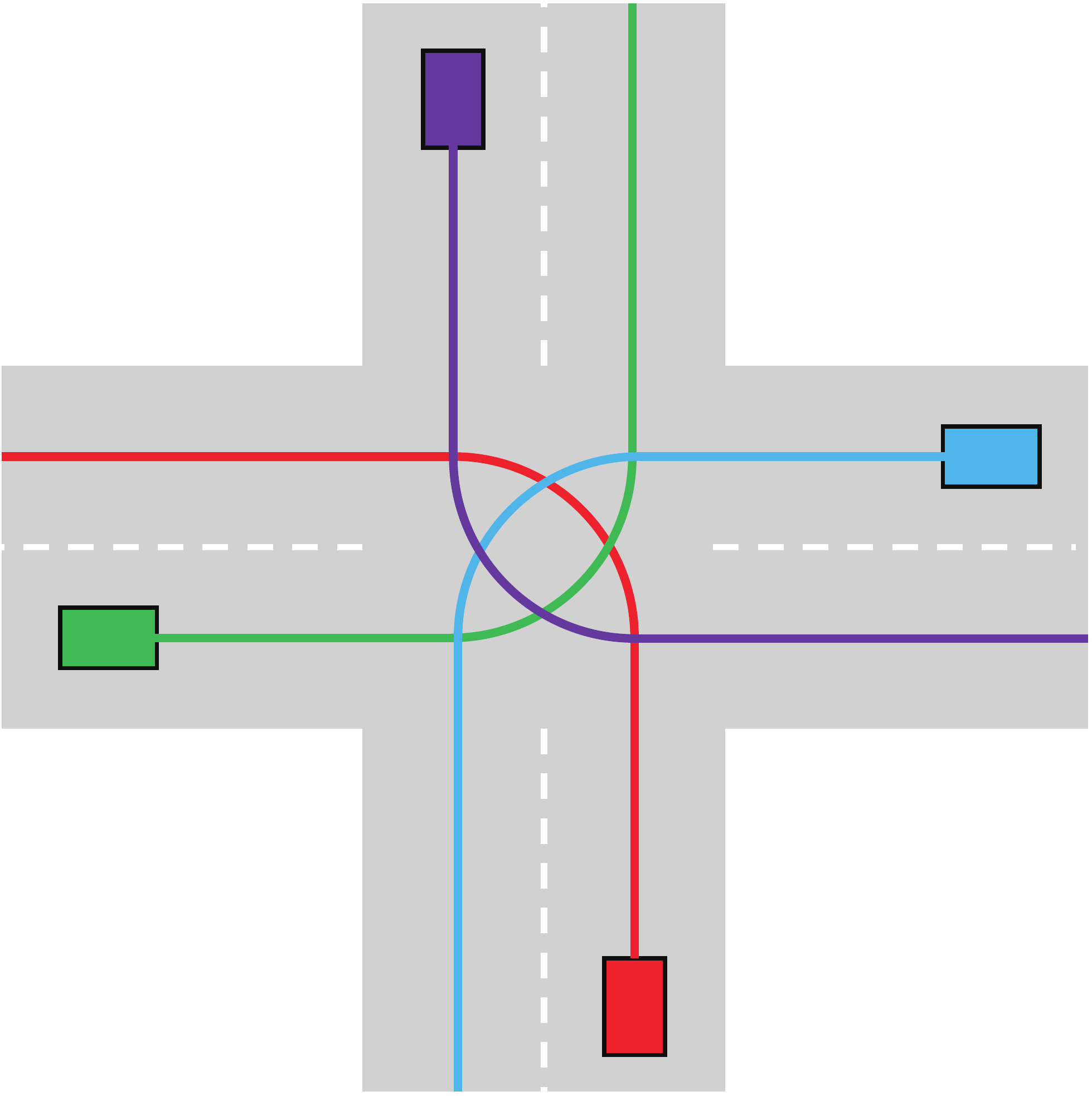}}
    \hfill
    \subfloat[space time resource searching]{%
        \includegraphics[scale=0.2]{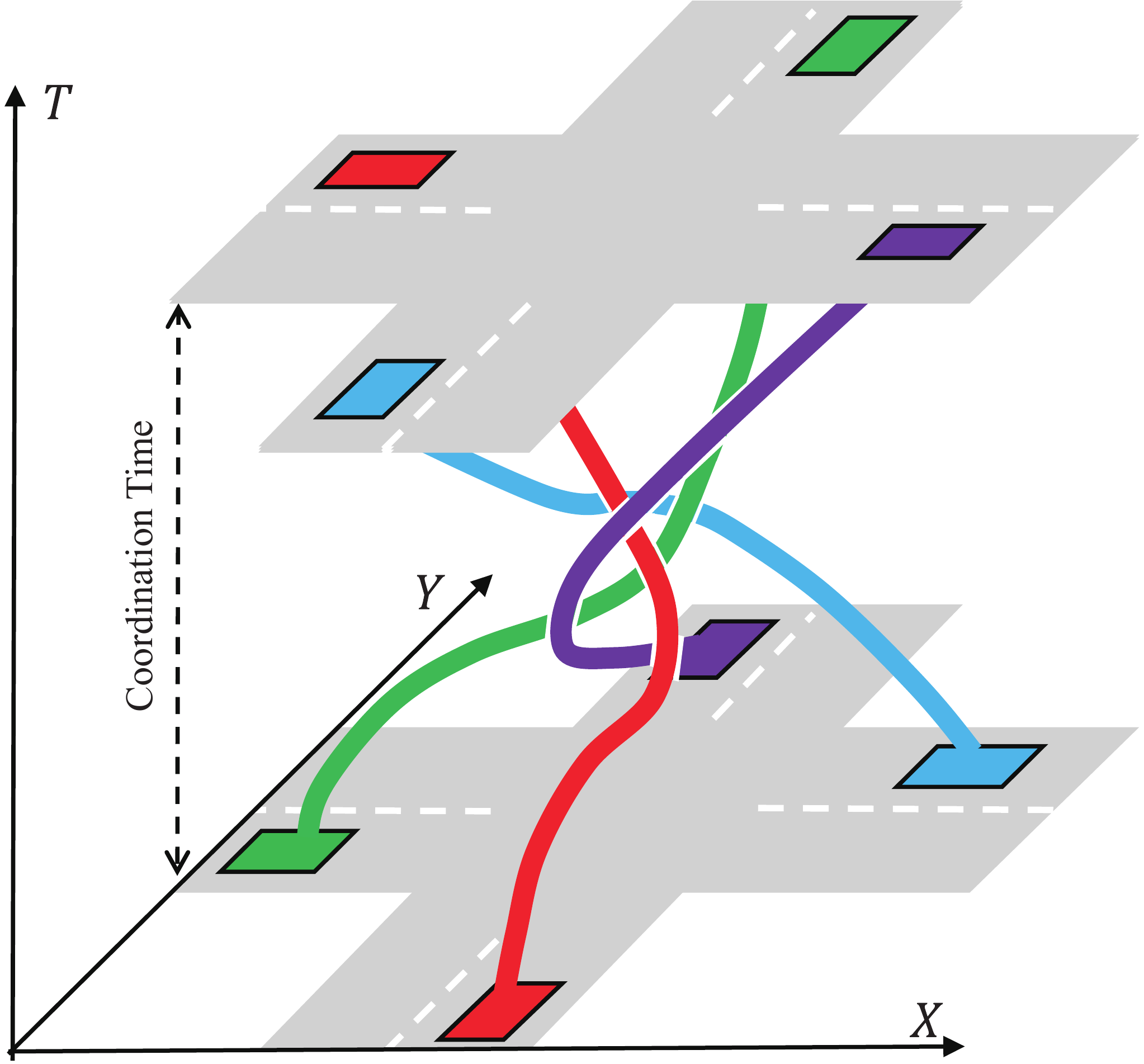}}
    \caption{An illustration of two benchmark strategies.}
    \label{fig:strategies}
\end{figure}

In the collision set (CS) strategy\cite{CS}, the area where collisions may occur is defined as the \textit{collision set}, where multiple vehicles cannot enter one collision set simultaneously.
For a typical single-lane scenario, as shown in Fig. \ref{fig:strategies}(a), any two left-turning vehicles may collide.
Therefore, the vehicles could only cross the intersection one after another. As a result, the collision-free constraints could be ignored and thus significantly reduce the complexity of solving the aforementioned problem $\mathscr{P}$.
Though the CS strategy shows superiority in terms of low-complexity implementation, it could not harness the full coordination potential of the intersection, especially for multi-lane scenarios.
\subsubsection{Space-Time Resource Searching}
The space-time resource searching (STRS) strategy\cite{STRS} adopts the aggressive idea that all CVs simultaneously share the intersection to improve traffic throughput.
Specifically, the space and time resources in the intersection can be regarded as three-dimensional coordinates with X, Y, and T domains, as shown in Fig. \ref{fig:strategies}(b).
The proposition of optimizing speed profiles to make CVs travel safely on the overlapped paths in the XY domain is equivalent to searching for disjoint trajectories in XYT domains.
Compared with the CS strategy, the STRS strategy shows obvious throughput advantages.
By traversing space-time resources, the STRS strategy could obtain optimal passing orders and trajectories, however, at the expense of real-time processing capabilities.
In practical implementations, the coordinator requires tremendous computational resources and power supply to deal with high computational complexity. Excessive problem-solving time could also lead to safety issues, especially in the CA. Moreover, the computational complexity grows exponentially with the number of vehicles and lanes, which will hinder this strategy for multi-lane scenarios.

\section{DRL-based Centralized Coordinator}\label{solution}

It is critical to rapidly obtain optimal trajectories for given initial configurations in the considered intersection coordination scenario.
However, searching for optimal trajectories in 3D space XYT, or optimizing the controls on given overlapped 2D paths, takes overlong problem-solving time.
Moreover, the computational complexity of the centralized coordination strategy grows exponentially with the number of lanes and the number of CVs.
Thus, it is challenging to adopt the pure optimization approach-based schemes to solve the optimal coordination problem in real-time.

To address the key computational challenges, an RL approach is leveraged in this paper. We first model the sequential decision-making problem $\mathscr{P}$ as a model-free MDP and then resort to DRL algorithms to tackle it.

\subsection{Problem Transformation}\label{problem_formulation}
We first formulate the intersection coordination problem using the reinforcement learning (RL) framework.
MDPs can be used to express the standard mathematical formalism of RL\cite{silver}. An MDP is a 5-tuple $(\mathnormal{S}, \mathnormal{A}, \mathnormal{R}, \mathnormal{P}, \gamma)$ that describes the interactions between the states of the environment and the actions of the RL agent. Specifically, at each time slot $t$, based on the observation $\mathbf{s}_t \in \mathnormal{S}$, an action $\mathbf{a}_t \in \mathnormal{A}$ sampled from a stochastic policy $\pi(\mathbf{a}_t|\mathbf{s}_t)$ or generated by a deterministic policy $\pi(\mathbf{s}_t)$ is executed by the RL agent and the environment transitions to next state $\mathbf{s}_{t+1} \in \mathnormal{S}$ with a transition probability $\mathrm{Pr}(\mathbf{s}_{t+1} | \mathbf{s}_t, \mathbf{a}_t) \in \mathnormal{P}$, receiving a reward $r_t(\mathbf{s}_t, \mathbf{a}_t) \in \mathnormal{R}$.
$\gamma \in [0, 1)$ is the discount factor that trades off long-term against short-term rewards.
According to whether the state transition probability and the expected return from any state-action pair are prior or not, RL algorithms can be classified into two categories: model-based and model-free. In the considered intersection coordination scenario, the agent has no idea how to make a proper decision for the subsequent time slot according to the current observed environment state. Hence, it can be regarded as a model-free one.

Deep RL (DRL), which combines RL with deep neural networks (DNNs), can be applied to deal with continuous control problems. By using DNNs as a non-linear function approximator over high-dimensional state spaces, only the network parameters need to be stored instead of a lookup table that records the values of all possible state-action pairs\cite{DQN}. Moreover, DNNs can infer actions on unseen observations, enabling the algorithm to obtain expected performance without traversing all state-action pairs.

\textit{1) Observation and Action Spaces}: Intuitively, the state of each CV should contain the position and velocity information. However, this information only portrays the current driving state and ignores the intention and association among CVs. For example, if the paths of the CVs do not overlap in the CA, they could leave the CCZ as quickly as possible. Otherwise, the CVs should collaboratively cross the CA to avoid collision in the subsequent time slots.

To capture the intentions of the CVs, a natural and straightforward scheme is to assign each vehicle a tag, i.e., 0-turn left, 1-go straight, and 2-turn right. However, DNNs have difficulty in identifying the meaning of these tags. To help the DNNs understand the tags, we could encode each of these tags as a three-element binary vector, e.g., [1, 0, 0] represents turning left. However, the required state dimensions to infer driving intentions increases linearly with the number of CVs, which is computationally inefficient for training the DRL agent. Finally, we use a $D_r$-element binary vector $\mathbf{d}_r$ to encode the joint intentions for CVs on the $r$-th road, where $D_r = 3^N$ if $K = 1$ else $D_r = 4^N$. This joint encode scheme shows effectiveness in identifying CVs' intentions using much fewer state dimensions.

Moreover, to capture the dynamic feature of the traffic, the queue state vector $\mathbf{I}(t)$ is also considered. We normalize the queue states to reflect the priority of different lanes, i.e.,
\begin{align}
    p_{r, k}(t) = \frac{I_{r, k}(t)}{\sum_{r \in \mathcal{R}}\sum_{k \in \mathcal{K}_r}I_{r, k}(t)} \label{eq:priority_normalize}
\end{align}
Then, the environment state at time slot $t$ can be described as:
\begin{align}
    \mathbf{s}_t = \{
             \mathbf{d}_r(t),  p_{r, k}(t), x_{r, k}^n(t), y_{r, k}^n(t), v_{r, k}^n(t) \notag \\
             | r \in \mathcal{R},
                          k \in \mathcal{K}_r, n \in \mathcal{N}_{r, k}
                   \} \label{eq:state}
\end{align}

The DRL agent is trained to directly map the observation to CVs' control commands, i.e., acceleration, velocity, heading angle, and steering angle. For a given reference path, the velocity, heading angle, and steering angle can be determined by acceleration. Hence, the agent only needs to output acceleration commands, which could greatly reduce the training complexity, thus
\begin{align}
    \mathbf{a}_t = \left\{a_{r, k}^n(t) | r \in \mathcal{R}, k \in \mathcal{K}_r, n \in \mathcal{N}_{r, k}\right\}
\end{align}

\begin{figure}
    \centering
    \includegraphics[scale=0.45]{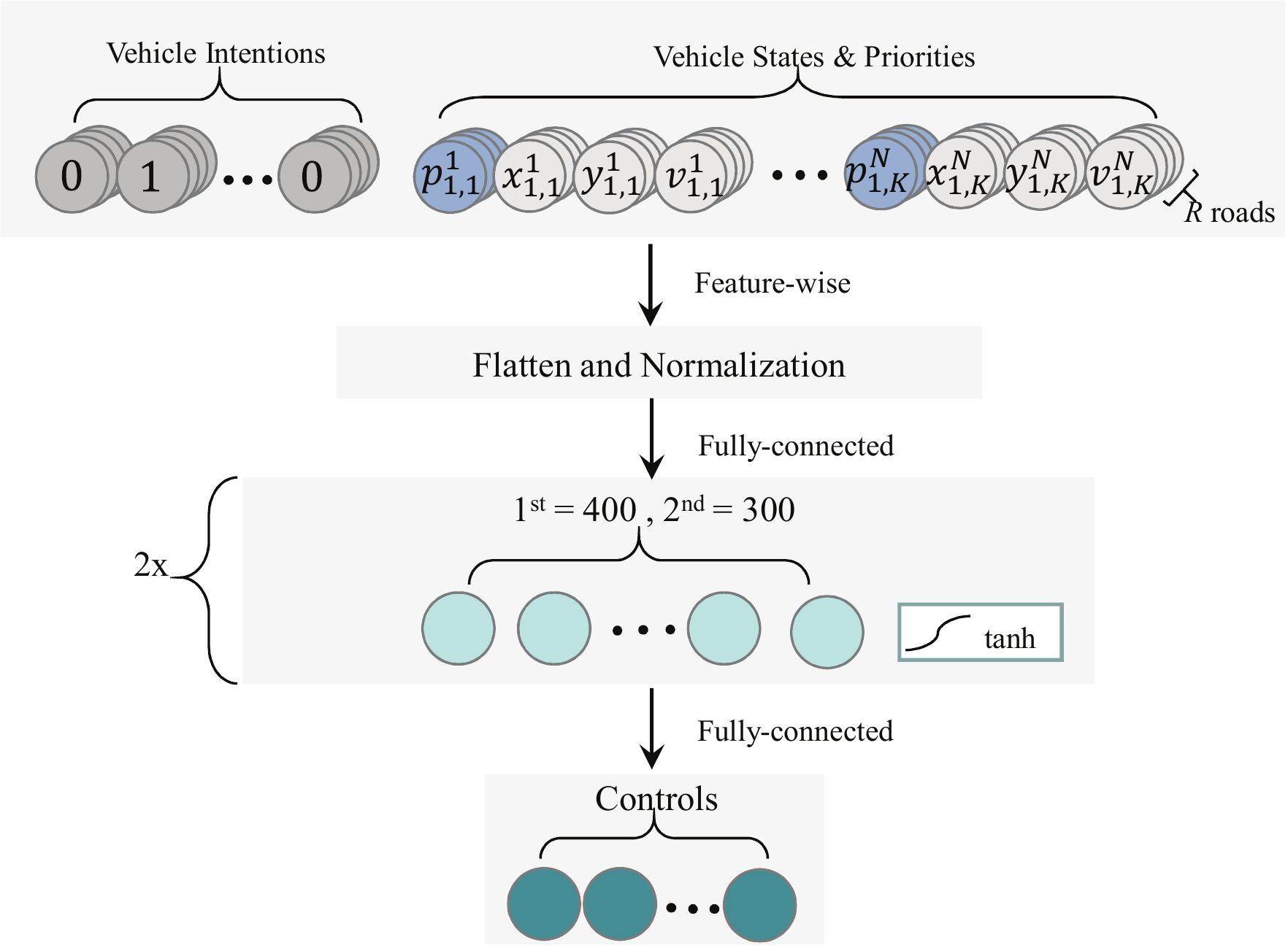}
    \caption{Illustration of the network architecture used in this work.}
    \label{fig:policy_net}
\end{figure}

Fig. \ref{fig:policy_net} shows the network architecture used in this work. The observation and action vectors are arranged according to the order of the roads. Note that the input and output layer of the policy network is fixed, so for the case where the number of CVs in lane $(r, k)$ is $N_{r, k}$ ($N_{r, k} < N$), we will add $N - N_{r, k}$ virtual vehicles and give them lower priority by initializing them at the edge of the CCZ. These virtual vehicles will not be used in calculating reward functions.

\textit{2) Reward Functions}: To some extent, the reward function determines how intelligent the RL agent would be. An unreasonable reward function may lead the agent to devise an ineffective or even unusable policy.

In general, the reward function should be designed based on the objective of the original optimization problem $\mathscr{P}$, i.e., $\sum_{r \in \mathcal{R}} \sum_{k \in \mathcal{K}_r} I_{r, k}(t)b_{r, k}(t)$. In order to reflect the priority of different lanes (or vehicles), we normalize the queue state $I_{r, k}(t)$ using \eqref{eq:priority_normalize}. Moreover, we use $T_{r, k}$ instead of $T_\text{batch}$ to motivate individual vehicles to leave the CA as soon as possible. The reward can then be written as $\sum_{r \in \mathcal{R}} \sum_{k \in \mathcal{K}_r} \frac{p_{r, k}(t) N_{r, k}(t)}{T_{r,k}}$. Given that $p_{r,k}$ already captures the queue state of lane $(r, k)$, term $N_{r,k}(t)$ can be omitted. Thus, the completion reward can be defined as:
\begin{align}
    r_\text{sa}(t) =
       10000 \mathbf{1}_t \cdot
       \left(
       \sum_{r \in \mathcal{R}} \sum_{k \in \mathcal{K}_r} \frac{p_{r, k}(t)}{T_{r,k}}
       \right)
\end{align}
where $\mathbf{1}_t$ is an indicator variable that represents whether all CVs (exclude virtual vehicles) leave the CCZ or not at time slot $t$, i.e., $\mathbf{1}_t = 1$ if all CVs leave CCZ and $\mathbf{1}_t = 0$ otherwise.

However, this reward can only be computed upon the completion of a full coordination, introducing sparsity that considerably increases the difficulty in evaluating the usefulness of individual actions. A popular approach to circumvent this problem is to use an immediate reward that roughly reflects the final goal while providing feedback to the agent at every time step\cite{drone_racing}. Therefore, we use velocity information to represent the effectiveness of individual actions, i.e.,
\begin{align}
    r_\text{v}(t) = \sum_{r\in \mathcal{R}} \sum_{k\in \mathcal{K}_r} \sum_{n\in \mathcal{N}_{r, k}}    v_{r, k}^n(t)
\end{align}

Most importantly, to learn a safe policy, we define a large negative reward to discourage the CVs from colliding in the CA, which can be expressed as:
\begin{align}
    r_\text{c}(t) = \begin{cases}
       -\, r_\text{T}, & \text{if collide} \\
       \,\,0,  & \text{otherwise}
    \end{cases}
\end{align}
where $r_\text{T}$ is a hyperparameter that trades off between safety and efficiency. Specifically, a large penalty may hinder the agent from exploring better decisions and result in a more conservative and inefficient policy. In contrast, a small penalty may lead to aggressive actions that compromise safety. In this paper, $r_\text{T}$ is set as 1000.

In addition, we define an intermediate reward to incentivize the vehicle to cross the intersection safely. It should be noted that this is an optional reward component designed to guide the agent to achieve our goal and improve the convergence speed and asymptotic performance of the DRL algorithm. The intermediate reward is defined as follows:
\begin{align}
    r_\text{s}(t) = 1000 n_t
\end{align}
where $n_t$ is the number of CVs that left the CCZ in time duration $[t, t+1)$. The final reward at each time slot $t$ could be defined as:
\begin{align}
    r_t = r_\text{sa}(t) + r_\text{v}(t) + r_\text{c}(t) + r_\text{s}(t) \label{eq:reward}
\end{align}

\subsection{TD3-based Solution}
The goal of the RL agent is to maximize the expected return by optimizing the policy over the environment dynamics\cite{SpinningUp2018,sutton}. The expected return $V$ under a policy $\pi$ can be calculated as:
\begin{align}
    V^{\pi}(\mathbf{s}_t) = \mathbb{E}_{\tau \sim \pi} \left[G_t \,| \, \mathbf{s}_t \right]
\end{align}
where $G_t = \sum_{i = t}^{\infty} \gamma^{i - t}r_i$ is the discounted return, $\tau = (\mathbf{s}_t, \mathbf{a}_t, \mathbf{s}_{t+1}, \mathbf{a}_{t+1}, \ldots)$ is the trajectory which the agent will follow from time step $t$, and $\tau \sim \pi$ means the agent acts according to the policy $\pi$. Value function (also called critic or Q-function) can be established to estimate $V^{\pi}$ in a bootstrapping way. Under an arbitrary policy $\pi$, value function can be defined as:
\begin{align}
    Q^{\pi}(\mathbf{s}_t, \mathbf{a}_t)
    &= \mathbb{E}_{\tau \sim \pi} \left[G_t \, |\, \mathbf{s}_t, \mathbf{a}_t\right] \notag \\
    &= r_t + \gamma \cdot \mathbb{E}_{\tau \sim \pi} \left[Q^{\pi}(\mathbf{s}_{t+1}, \mathbf{a}_{t+1}) \right] \label{eq:Q}
\end{align}
Then the policy function (also called actor) can be obtained by maximizing value function, represented as:
\begin{align}
    \pi^* = \arg\max_{\pi} \mathbb{E}_{\tau \sim \pi} \left[Q^{\pi}(\mathbf{s}_t, \mathbf{a}_t)\right]
    \label{eq:pi}
\end{align}
Equations (\ref{eq:Q}) and (\ref{eq:pi}) concrete the actor-critic RL framework. In practical implementations, we utilize neural networks as the differentiable function approximator to construct the actor and critic. Hence, the optimal policy can be obtained by optimizing the parameters of the policy network $\pi_{\mathbf{\theta}}$ and value network $Q_{\mathbf{\phi}}$.

\begin{figure}[htp]
    \centering
    \includegraphics[scale=0.35]{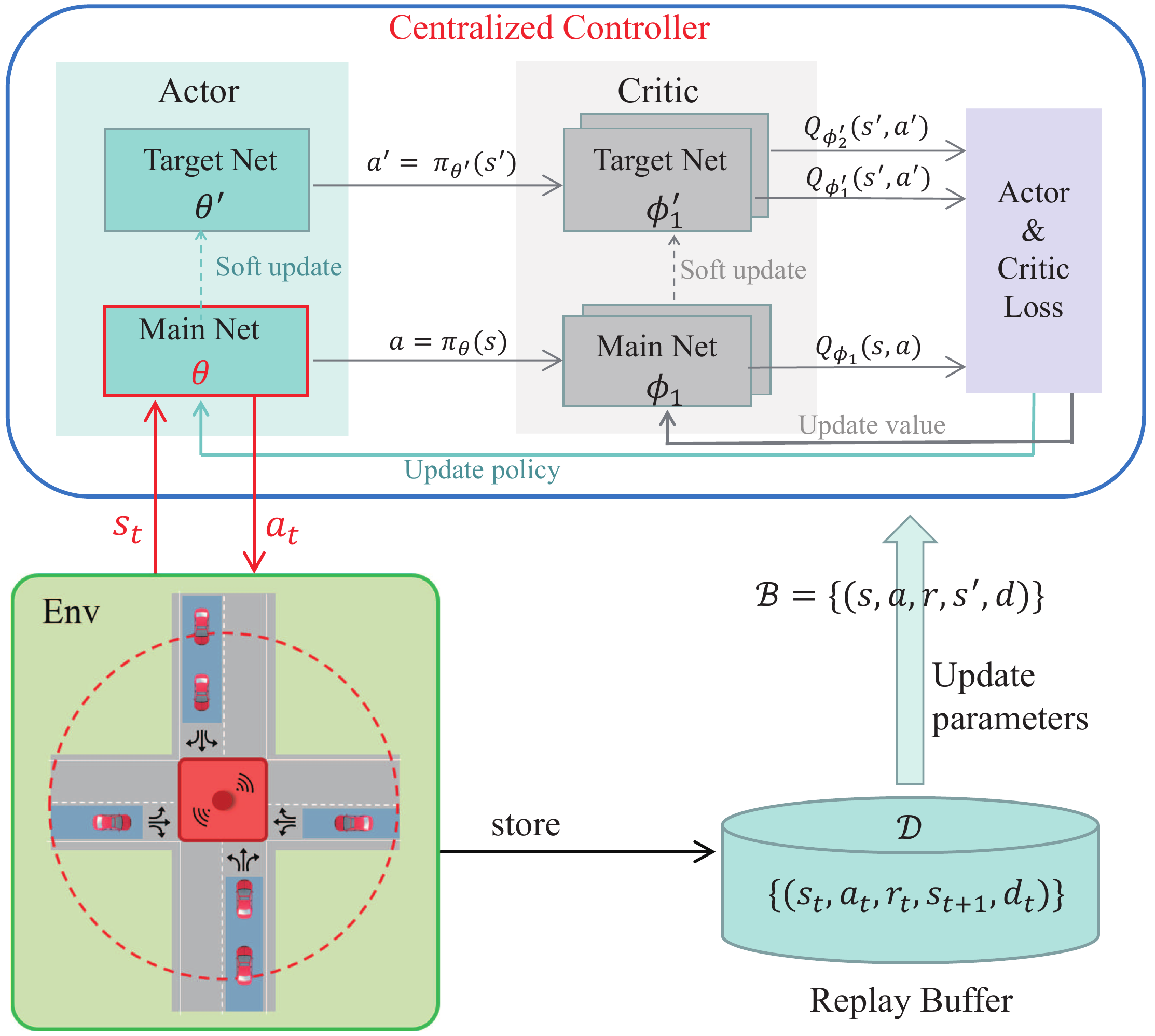}
    \caption{The architecture and training process of TD3 agent.}
    \label{fig:TD3}
\end{figure}

In this work, we train our agent using the TD3 algorithm\cite{TD3}, a state-of-the-art actor-critic method that is particularly popular in continuous control problems such as robotics and autonomous driving.
As shown in Fig. \ref{fig:TD3}, the TD3 agent consists of six neural networks of two types: the main networks and the target networks, and each type contains one policy network and two value networks. The target and main networks share the same structure but different parameters.
At time step $t$, each CV shares its driving status with the agent, and based on the observation $\mathbf{s}_t$, the main policy network $\pi_{\mathbf{\theta}}$ outputs the control commands for the CVs, i.e.,
\begin{align}
    \mathbf{a}_t = \mathop{\mathrm{clip}}( \pi_{\mathbf{\theta}}(\mathbf{s}_t) + \epsilon, -a_\text{max}, a_\text{max}) \label{eq:action}
\end{align}
where $\epsilon \sim \mathcal{N}(0, \sigma)$ is the exploration noise aiming to encourage the agent to take the risk of trying new actions which may lead to higher rewards, rather than exploiting its knowledge by selecting actions which are known to result in high rewards.
By executing $\mathbf{a}_t$, the vehicles transition to the next state $\mathbf{s}_{t+1}$ according to the vehicle kinematic model (\ref{eq:dynamic}) and receive a reward $r_t$ which evaluates the usefulness of action $\mathbf{a}_t$.
For more efficient use of previous experience and better convergence performance when training the TD3 agent, a replay buffer $\mathcal{D}$ is considered to save each experience $(\mathbf{s}_t, \mathbf{a}_t, r_t, \mathbf{s}_{t+1}, d_t)$, where $d_t$ indicates whether the state $\mathbf{s}_{t+1}$ is terminal (terminate on a collision or successful coordination).

During the training stage, a mini batch of experiences $\mathcal{B} = \{(\mathbf{s}, \mathbf{a}, r, \mathbf{s}^\prime, d)\}$ are randomly sampled from the replay buffer and then fed to the agent in parallel to update the neural networks. The main value networks can be updated by minimizing the following critic loss function:
\begin{align}
    \mathcal{L}^Q(\mathbf{\phi}) = \frac{1}{|\mathcal{B}|} \sum_{(\mathbf{s}, \mathbf{a}, r, \mathbf{s}^\prime, d) \in \mathcal{B}}
    \left(Q_{\mathbf{\phi}}(\mathbf{s}, \mathbf{a}) - y(r, \mathbf{s}^\prime , d)\right)^2
\end{align}
where $y(r, \mathbf{s}^\prime, d)$ is called the temporal difference (TD)-target, which represents the latest prediction of the Q-function.
In an actor-critic setting, the TD-target is usually formulated as $r + \gamma (1-d) Q_{\phi^{\prime}}(\mathbf{s}^\prime, \pi_{\theta^{\prime}}(\mathbf{s}^\prime))$, where $Q_{\phi^{\prime}}$ and $\pi_{\theta^{\prime}}$ are the target value network and target policy network\cite{DDPG}, respectively.
In practice, however, using one target value network to update the estimate $Q_{\mathbf{\phi}} (\mathbf{s}, \mathbf{a})$ will lead to overestimation bias, which could result in suboptimal policy updates and divergent behavior. In TD3, a novel method, Clipped Double Q-learning, is designed to greatly reduce overestimation bias. The TD-target in the TD3 algorithm is defined as:
\begin{align}
    y(r, \mathbf{s}^\prime, d) = r + \gamma (1-d) \min_{i = 1, 2}Q_{\phi_{i}^\prime}(\mathbf{s}^\prime, \pi_{\theta^\prime}(\mathbf{s}^\prime) + \Tilde{\epsilon}) \label{eq:TD}
\end{align}
where $Q_{\mathbf{\phi}^{\prime}_1}$,  $Q_{\mathbf{\phi}^{\prime}_2}$ are the two target value networks, $\Tilde{\epsilon} \sim \mathop{\mathrm{clip}}(\mathcal{N}(0, \Tilde{\sigma}), -c, c)$ is the clipped policy noise, $c \in (0, 1)$ is the clip ratio. Then, the main value networks can be updated by gradient descent using
\begin{align}
    \nabla_{\phi_i}\mathcal{L}^Q(\phi_i) = \nabla_{\phi_i}
    \frac{1}{|\mathcal{B}|} \sum_{(\mathbf{s}, \mathbf{a}, r, \mathbf{s}^\prime, d) \in \mathcal{B}}
    \left(Q_{\phi_i}(\mathbf{s}, \mathbf{a}) - y(r, \mathbf{s}^\prime , d)\right)^2 \label{eq:critic}
\end{align}
The policy network is updated by maximizing Q-function through deterministic policy gradient algorithm\cite{DPG}:
\begin{align}
    \nabla_{\theta}\mathcal{L}^{\pi}(\theta)
    &= \nabla_{\theta} \frac{1}{|\mathcal{B}|}
    \sum_{\mathbf{s} \in \mathcal{B}}
    Q_{\phi_1}(\mathbf{s}, \pi_{\theta}(\mathbf{s})) \notag \\
    &= \frac{1}{|\mathcal{B}|}
    \sum_{\mathbf{s} \in \mathcal{B}}
    \left(
    \nabla_a Q_{\phi_1}(\mathbf{s}, \mathbf{a})\big|_{\mathbf{a} = \pi_{\theta}(\mathbf{s})}
    \nabla_{\theta}\pi_{\theta}(\mathbf{s})
    \right) \label{eq:actor}
\end{align}

Considering the overestimation bias caused by value networks, the policy network and target networks should be updated at a lower frequency than the value network. So we only update the policy and target networks after a fixed number of updates to the critic. The training stage of the TD3-based centralized coordinator is summarized in \textbf{Algorithm \ref{algorithm_1}}.

The policy network $\pi_\theta$ trained offline can be regarded as a real-time solver, which could instantly output current controls by inputting an observation of the environment.
However, interacting with the environment at each time slot will waste tremendous communication resources, and the communication delay also causes jitter in control.
In practical implementations, the coordinator can operate in \textit{predictive mode}: all instructions for the subsequent time slots are sent to the CVs at once by simulating vehicle kinematics. It is reasonable when the noise in control, sensing and localization is relatively small. Otherwise, a re-planning operation will be necessary when the trajectory deviation is unbearable.
\begin{algorithm}[htb]
    \caption{Training Stage of The TD3-based Coordinator}
    \label{algorithm_1}
    \begin{algorithmic}[1]
        \STATE Initialize policy parameters $\theta$, value parameters $\phi_1$, $\phi_2$. Empty replay buffer $\mathcal{D}$. Set target parameters equal to main parameters $\theta^\prime \longleftarrow \theta$, $\phi_1^\prime \longleftarrow \phi_1$, $\phi_2^\prime \longleftarrow \phi_2$. The parameters are updated after $T_\text{start}$ steps, the policy network and target networks are updated after $T_\text{delay}$ updates of the critic. Initialize total time steps $T_\text{total} = 0$, total network updates $T_\text{updates} = 0$.\\
        \FOR{$episode = 1\,  \text{to}\,  E$}
            \STATE Randomly initialize the position of the CVs and the number of WVs, receive initial observation $\mathbf{s}_0$. Set time step $t = 0$, done signal $d_0 = \text{False}$.\\
            \WHILE{not $d_t$}
                \STATE Observe current environment state $\mathbf{s}_t$.
                \IF{$T_\text{total} < T_\text{start}$}
                      \STATE Randomly sample an action from action space.
                \ELSE
                      \STATE Select action using policy $\pi_{\theta}$ according to (\ref{eq:action}).
                \ENDIF
                \STATE The CVs execute control commands $\mathbf{a}_t$ according to vehicle kinematic model (\ref{eq:dynamic}).\\
                \STATE Check collision using SAT algorithm. \\
                \IF{collide or complete}
                      \STATE Set done signal $d_t$ = True.\\
                \ENDIF
                \STATE Observe next state $\mathbf{s}_{t+1}$, compute reward $r_t$ according to (\ref{eq:reward}), check done signal $d_t$.\\
                \STATE Store experience $(\mathbf{s}_t, \mathbf{a}_t, r_t, \mathbf{s}_{t+1}, d_t)$ in $\mathcal{D}$.\\
                \IF{$T_\text{total} > T_\text{start}$}
                      \STATE $T_\text{updates} \longleftarrow T_\text{updates} + 1$\\
                      \STATE Randomly sample a mini-batch of transitions $\mathcal{B} = \{(\mathbf{s}, \mathbf{a}, r, \mathbf{s}^\prime, d)\}$ from $\mathcal{D}$.\\
                      \STATE Compute TD-targets according to (\ref{eq:TD}).\\
                      \STATE Update $\phi_1$ and $\phi_2$ by gradient descent using (\ref{eq:critic}).\\
                      \IF{$ T_\text{updates}\, \, \text{mod} \, \, T_\text{delay} = 0$}
                           \STATE Update $\theta$ by gradient ascent using (\ref{eq:actor}).\\
                           \STATE Update target networks using \textit{soft update}:
                           \begin{align*}
                               \theta^\prime &\longleftarrow \tau \theta + (1 - \tau) \theta^\prime \\
                               \phi_i^\prime &\longleftarrow \tau \phi_i + (1 - \tau) \phi_i^\prime, \, \text{for}\, i = 1, 2
                           \end{align*}\\
                      \ENDIF
                \ENDIF
                \STATE $T_\text{total} \longleftarrow T_\text{total} + 1$\\
            \ENDWHILE
        \ENDFOR
    \end{algorithmic}
\end{algorithm}

\subsection{Dynamic Coordination Framework}
We observe that the original trajectories generated by the TD3 agent are significantly jittery before entering the CA due to the lack of passenger comfort metrics in the reward function. To this end, we further smooth the original trajectories in \textit{fr\'{e}net frame}\cite{frenet}. For a given start state $[s_0, v_0, a_0]$ at $t_0$ and an end state $[s_1, v_1, a_1]$ at $t_1 = t_0 + T$ ($t_1$ is the time the vehicle enters the CA),
we generate a smooth longitudinal trajectory using quintic polynomials,
\begin{align}
    s(t) =    \omega_0 + \omega_1 t   + \omega_2 t^2
         + \omega_3t^3 + \omega_4 t^4 + \omega_5 t^5 \label{eq:smoother}
\end{align}
Here, the coefficients can be calculated by solving the following linear function:
\begin{align}
    \begin{bmatrix}
        s_0\\
        v_0\\
        a_0\\
        s_1\\
        v_1\\
        a_1
    \end{bmatrix}
    =
    \begin{bmatrix}
        1 & 0 & 0 & 0 & 0 & 0 \\
        0 & 1 & 0 & 0 & 0 & 0 \\
        0 & 0 & 2 & 0 & 0 & 0 \\
        1 & T & T^2 & T^3 & T^4 & T^5\\
        0 & 1 & 2T  & 3T^2& 4T^3 & 5T^4\\
        0 & 0 & 2   & 6 T & 12 T^2 & 20T^3
    \end{bmatrix}
    \cdot
    \begin{bmatrix}
        \omega_0\\
        \omega_1\\
        \omega_2\\
        \omega_3\\
        \omega_4\\
        \omega_5
    \end{bmatrix}\label{eq:coefficients}
\end{align}

As discussed in Section \ref{problem_formulation}, to cope with unbalanced arrival rates and demands in dynamic coordination scenario, vehicle grouping and vehicle padding methods are considered. Specifically, WVs need to be divided into batches for the lane with more than $N$ WVs, and virtual vehicles will be added for the lane with less than $N$ WVs. For each round, the TD3 agent is responsible for giving instructions to the CVs and ensuring driving safety.
In general, the proposed dynamic coordination framework is summarized in \textbf{Algorithm \ref{algorithm_2}}.

\begin{algorithm}[htb]
    \caption{Dynamic Coordination Framework}
    \label{algorithm_2}
    \begin{algorithmic}[1]
        \STATE The coordinator load pre-trained policy network $\pi_{\theta}$.
        \REPEAT
        \STATE Send Velocity Keeping and Safe-distance Maintaining commands to all WVs\cite{frenet}. Divide WVs into batches and add virtual vehicles if needed. Current batch of CVs hand over their control authorities to the coordinator.
        \STATE Collect all state information including queue states, CV's driving states and intentions.
        \STATE Construct initial state vector according to (\ref{eq:state}).
        \STATE The coordinator operate in predictive mode and compute optimal trajectories for all CVs using policy $\pi_{\theta}$.
        \STATE Smooth the original trajectories using (\ref{eq:smoother}) and (\ref{eq:coefficients}).
        \REPEAT
            \STATE Current batch of CVs cross the intersection according to maneuver instructions.
            \IF{CV $(r, k, n)$ leave the CA}
                 \STATE Turn green and keep a velocity $v_\text{max}$.
            \ENDIF
        \UNTIL{All CVs leave the CA (exclude virtual vehicles).}
        \STATE Completion of current coordination. Transition to next batch of CVs.
        \UNTIL{The coordination process is finished.}
    \end{algorithmic}
\end{algorithm}


\section{Simulation Results and Experiments}\label{simulation_results}

\subsection{Simulation Settings}
To demonstrate the performance of the proposed TD3-based coordination strategy, simulation results and experiments are presented in this section. To facilitate comparison with the strategies described in Section \ref{sec:benchmarks}, we consider a special case with batch size $N$ = 1. In Section \ref{batch_size}, we will further discuss the influence of different batch sizes on the coordination performance. The simulation procedure can be summarized in two stages, i.e., the training stage to optimize network parameters and the evaluation stage to test the trained agent.
\begin{figure}[htp]
    \centering
    \includegraphics[scale=0.3]{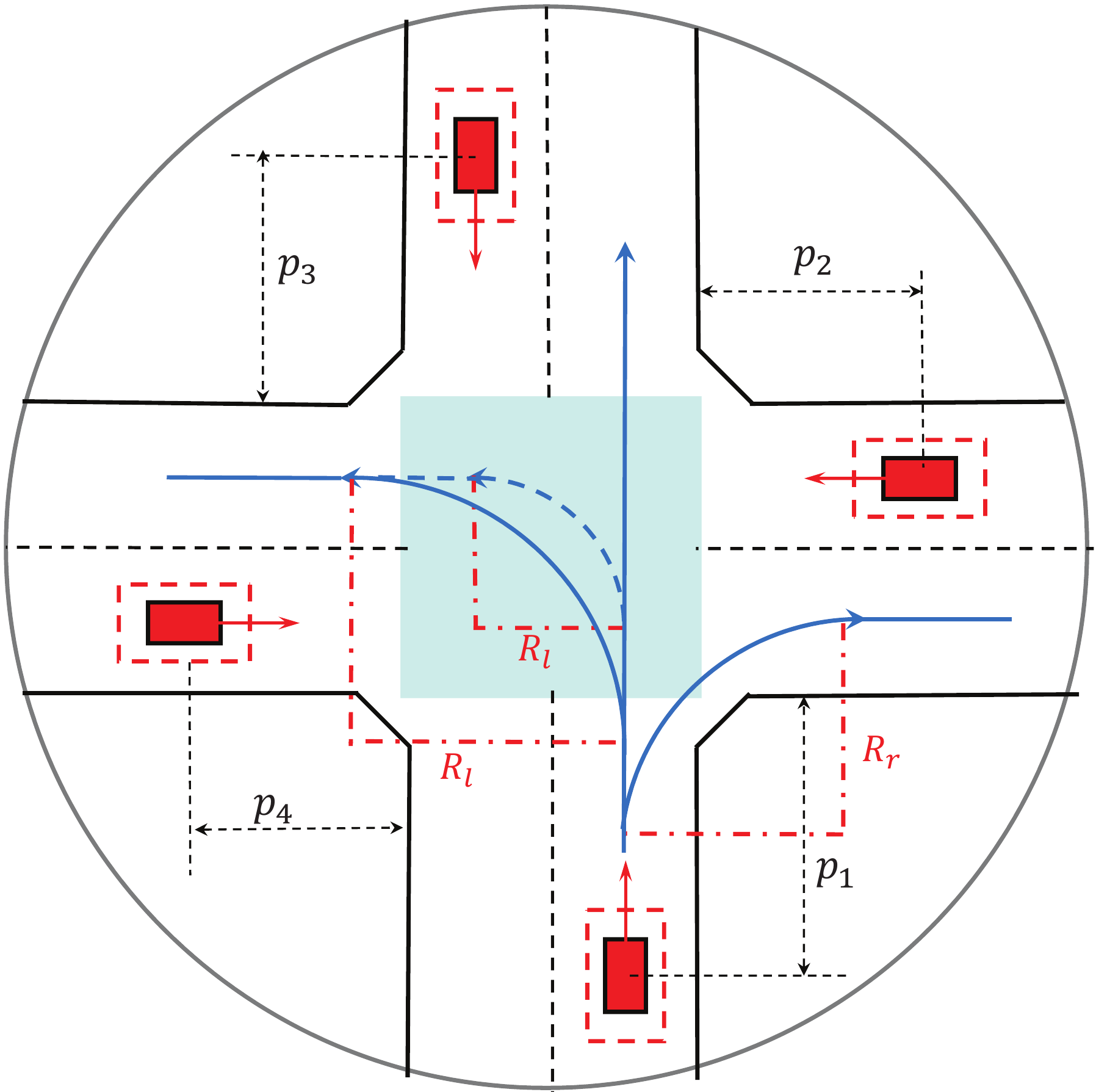}
    \caption{An illustration of the single-lane scenario with different left-turning radii.}
    \label{fig:single_lane}
\end{figure}

In the training stage, we first train separate policies for single-lane and multi-lane scenarios, as shown in Fig. \ref{fig:single_lane} and Fig. \ref{fig:intersection}, respectively. Without loss of generality, the position, velocity and intention of the vehicles are randomly initialized, and the paths of the left/right-turning vehicles are line \& circle curves. The left/right-turning radius, denoted by $R_\text{l}/R_\text{r}$, is set as 15(10)m/15m for single-lane case and 40m/20m for multi-lane case. Therefore, three separate policies are trained: (1) single-lane road intersection with $R_\text{l} = 15\text{m}$. (2) single-lane road intersection with with $R_\text{l} = 10\text{m}$. (3) multi-lane road intersection. The training process confronts challenges due to the high dimensionality of searching space, the complexity of required maneuvers, and the randomness of initial states.
To this end, we adopt the \textit{parallel sampling scheme}, which allows for increasing the number of agents to interact with different environments in parallel, resulting in a significant speed-up in data collection. Moreover, parallelization is an effective way to enrich the diversity of the collected experiences, which is beneficial to avoiding policy overfitting and increasing the probability of successful coordination. Detailed training parameters of the single-lane and multi-lane scenarios are listed in Table \ref{tab:table_1}.

In the evaluation stage, batchmark strategies discussed in Section \ref{sec:benchmarks} and the tradition TSC system are also performed for comparison. For ease of analysis, both static and dynamic coordination tasks are considered, where static coordination refers to the coordination of a single batch of CVs, while dynamic coordination refers to the coordination of realistic traffic flow.
There are four kinds of performance indices evaluated in the simulations:

\begin{itemize}
    \item \textit{total passing time}: it refers to the total time when a single batch of CVs pass the CCZ. This metric can be used to compare the control efficiency in static coordination scenarios.
    \item \textit{average computation time}: it refers to the average computation time taken for coordinating a single batch of CVs, which can be used to evaluate the computational efficiency of different strategies.
    \item \textit{average travel time}: it refers to the average time a vehicle spends from entering to exiting the CCZ. This metric is used to test the coordination performance in continuous traffic flow.
    \item \textit{traffic throughput}: it refers to the average passing time for a specified number of WVs queued up in the CCZ. This metric can be utilized to evaluate the maximum coordination rate of a certain strategy.
\end{itemize}

\begin{table}[t]
    \centering
    \caption{Simulation Parameters}
    \label{tab:table_1}
    \begin{tabular}{ccc}
        \toprule
   \multirow{2}{*}{\textbf{Parameters}}&     \multicolumn{2}{c}{\textbf{Value}}         \\
                                              \cline{2-3}
                             &              Single-lane             &    Multi-lane           \\
         \midrule
        Number of lanes      &         1                            &     3                   \\
        Radius of CCZ        &      50 m                            &    100 m                \\
        Size of CA           &       20 m $\times$ 20 m             &      48 m $\times$ 48 m \\
 left/right-turning radius   &         15(10) m / 15 m              &        40 m / 20 m      \\
Vehicle size $L_\text{car}\,/\,W_\text{car}$ &        4 m / 2 m     &    4 m / 2 m            \\
Safe redundancy $d_\text{lon}\,/\,d_\text{lat}$ &        4 m / 2 m  &    4 m / 2 m            \\
Maximum velocity $v_\text{max}$&     15 m/s                         &           15 m/s        \\
Maximum acceleration $a_\text{max}$ &     5 $\text{m/}\text{s}^2$   & 5 $\text{m/}\text{s}^2$ \\
Maximum steering angle $\delta_\text{max}$ &    0.78 rad            &      0.78 rad           \\
WV's velocity $v_\text{low}$ &        5 m/s                         & 5 m/s                 \\
Slot duration $\Delta T$     &           0.1 s                      &      0.1 s              \\
Number of parallel agents    &            6                         &          12              \\
      Network architecture   &    $[28, 256, 256, 4]$               &    $[64, 400, 300, 12]$ \\
Learning rate of actor/critic&     3e-4 / 3e-4                      &   3e-4 / 3e-4           \\
Soft update rate $\tau$      &       5e-3                           &  5e-3                   \\
 Exploration noise $\sigma$  &        0.1                           &     0.1                 \\
 Policy noise $\Tilde{\sigma}$ / clip ratio $c$&       0.2 / 0.5  & 0.2 / 0.5               \\
 Discount factor $\gamma$    &         0.99                         & 0.99                    \\
 Mini-banch size $|\mathcal{B}|$&          128                      &   256                   \\
 Maximum episodes $E$        &            2000                       &      20000               \\
 Training delay $T_\text{start}$&          25e3                     &     50e3                \\
 Policy delay $T_\text{delay}$  &           2                       &         2               \\
        \bottomrule
    \end{tabular}
\end{table}

\subsection{Single-lane Scenario}
\begin{figure}[htp]
    \centering
    \includegraphics[scale=0.45]{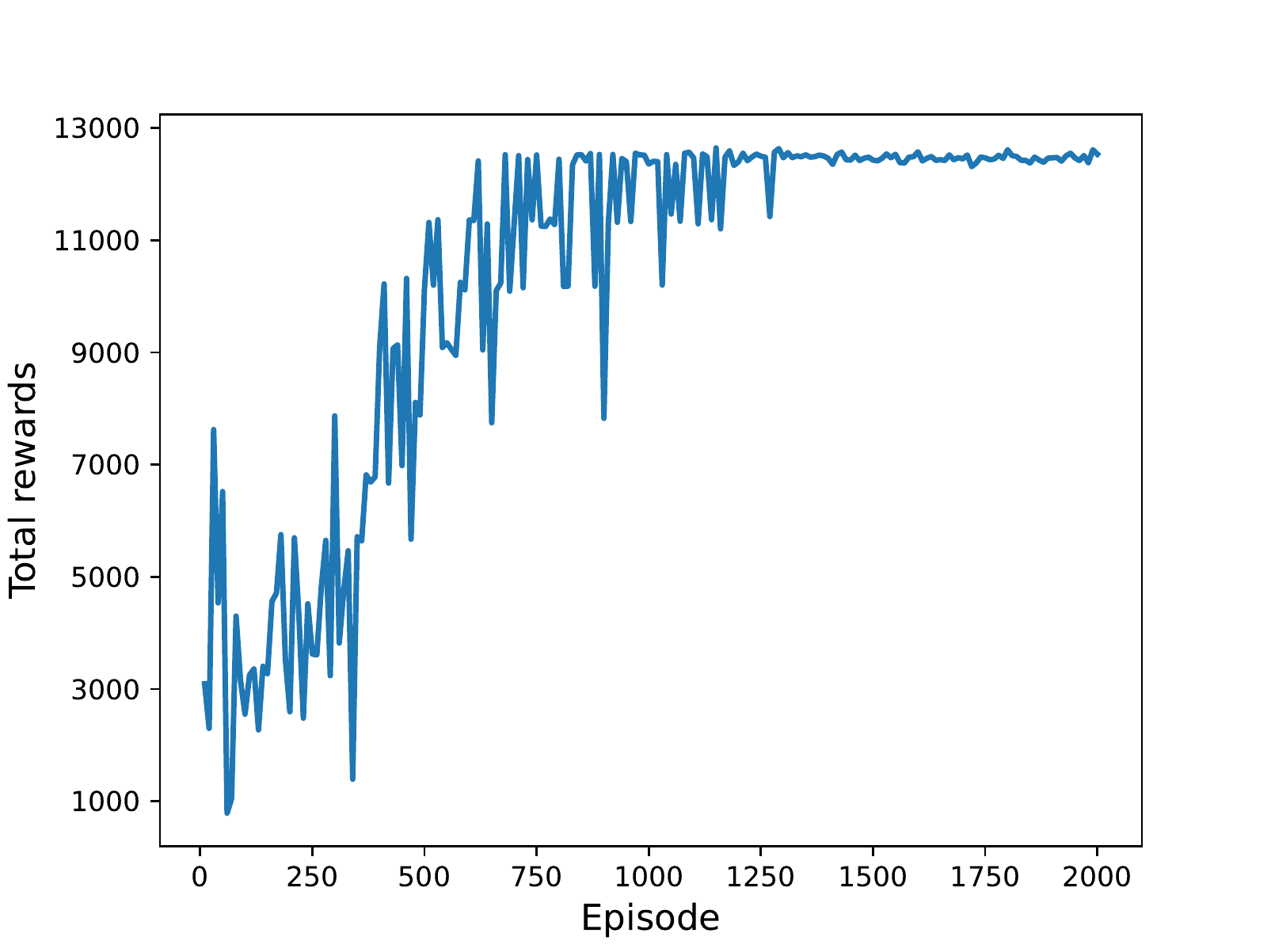}
    \caption{Convergence performance of the proposed TD3-based strategy.}
    \label{fig:convergence_performance}
\end{figure}
\begin{figure*}[htp]
    \centering
    \subfloat[acceleration]{%
        \includegraphics[scale=0.295]{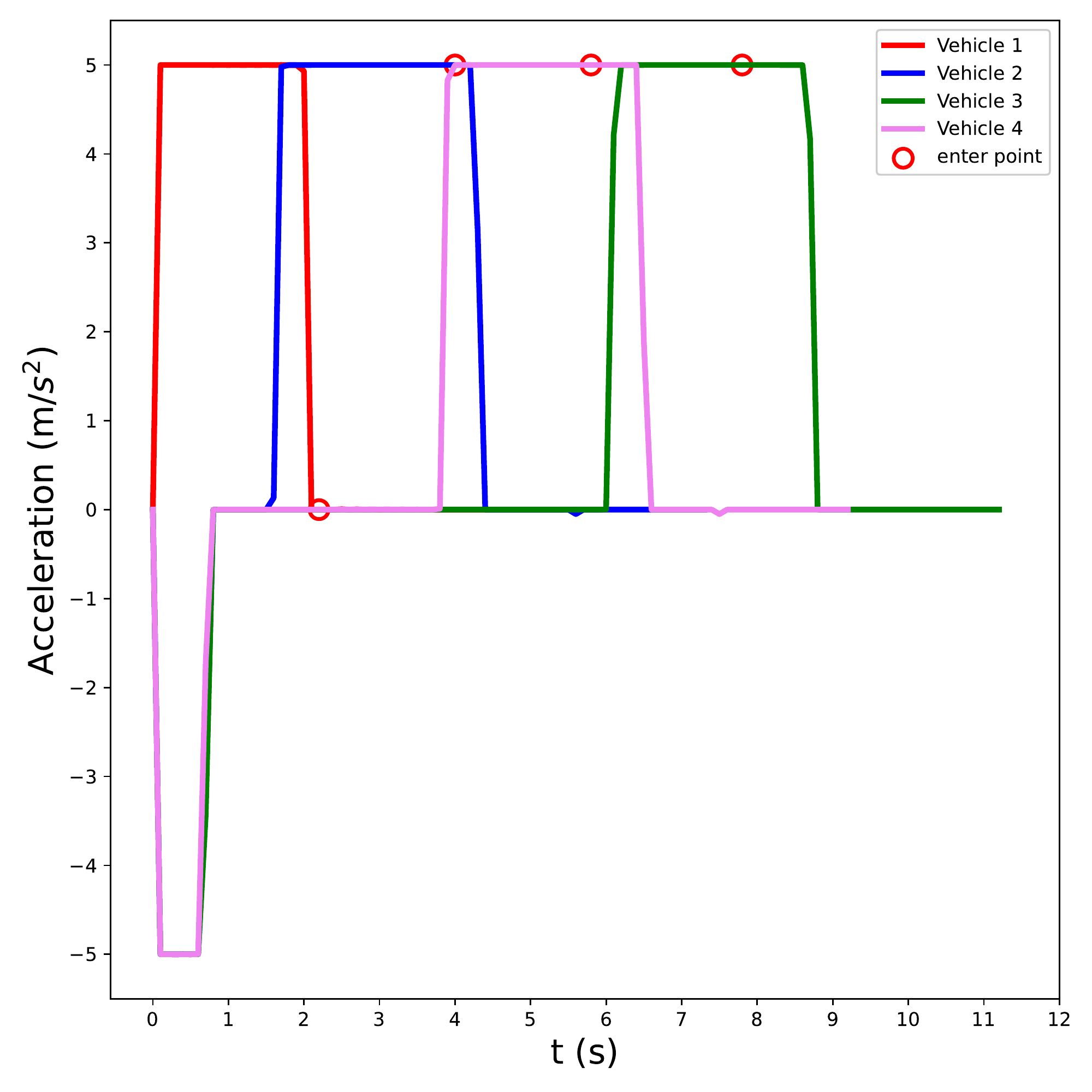}}
    \subfloat[velocity]{%
        \includegraphics[scale=0.295]{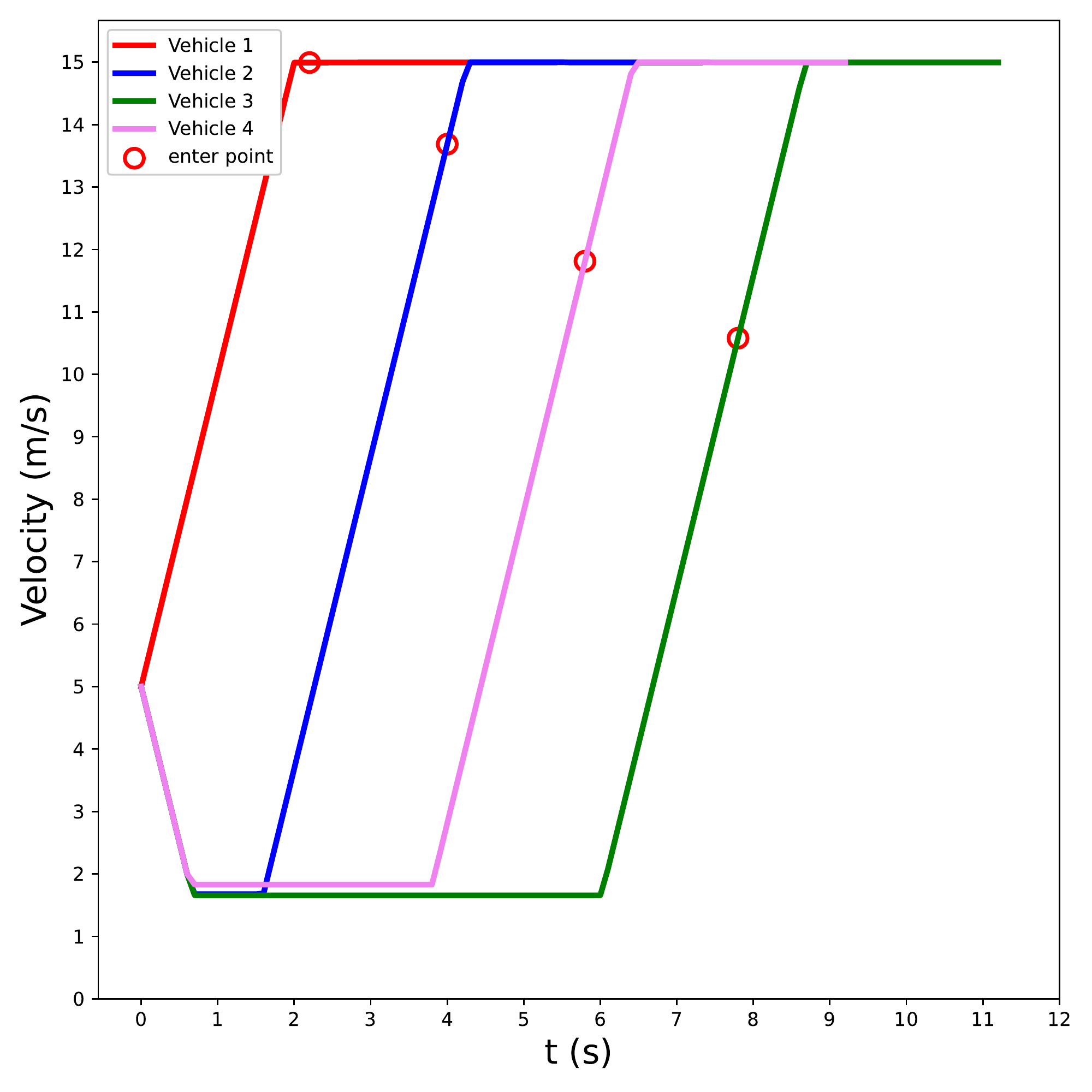}}
    \subfloat[longitudinal distance]{%
        \includegraphics[scale=0.295]{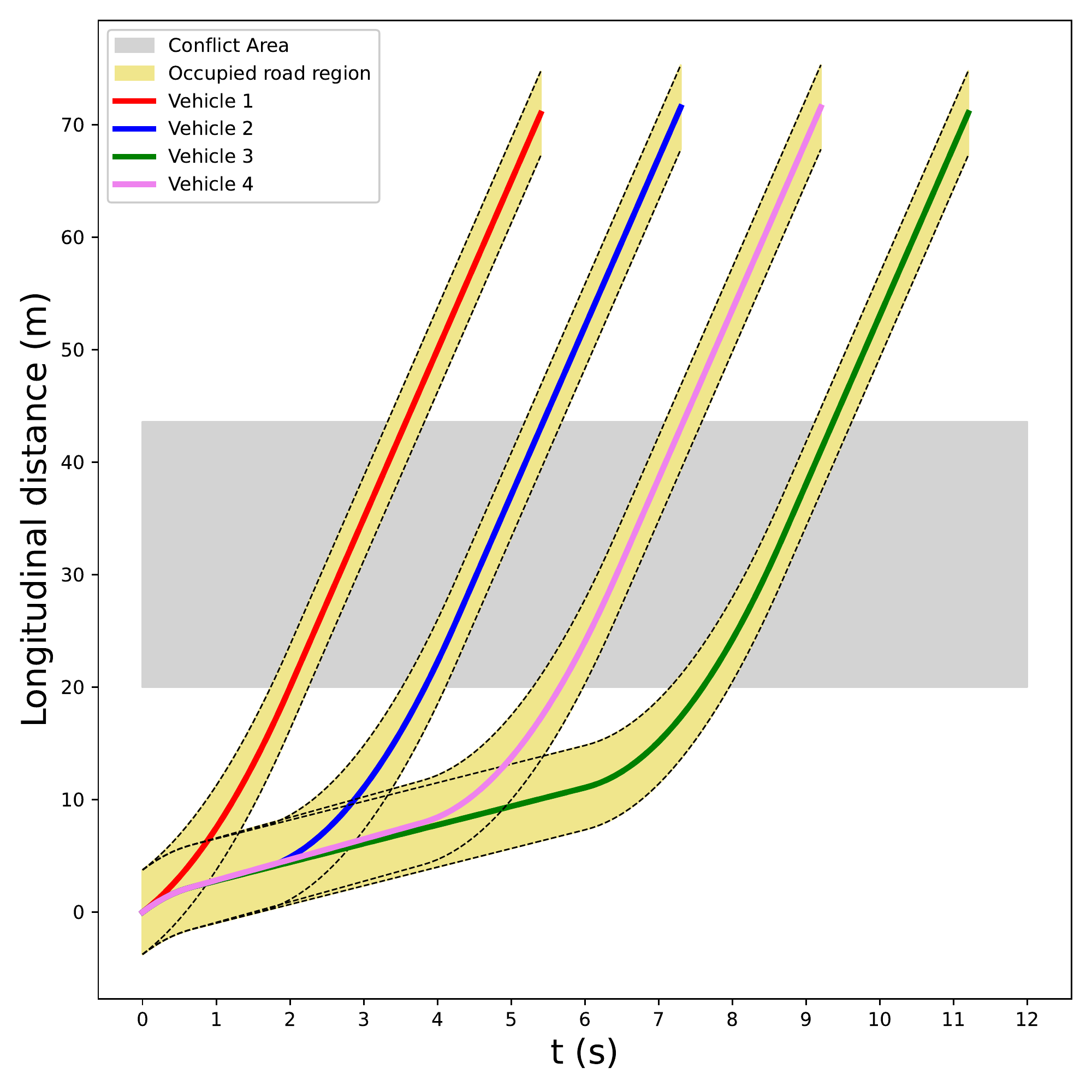}}
    \caption{The optimized outputs of the CS-based coordinator ($R_\text{l} = 15\text{m}$).}
    \label{fig:CS_acc_vel_ST}
\end{figure*}
\begin{figure*}[htp]
    \centering
    \subfloat[acceleration\label{fig:acc}]{%
        \includegraphics[scale=0.295]{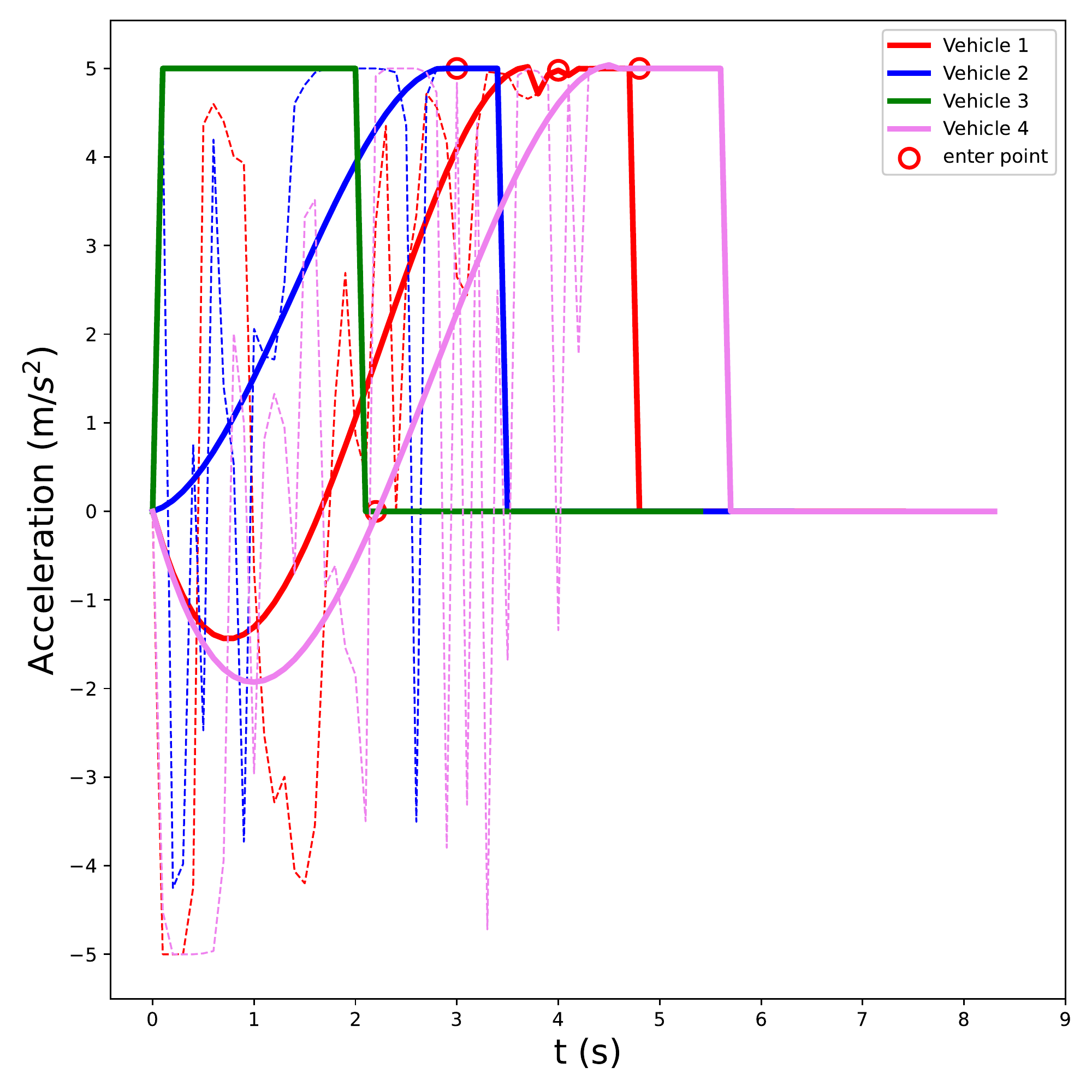}}
    \subfloat[velocity\label{fig:vel}]{%
        \includegraphics[scale=0.295]{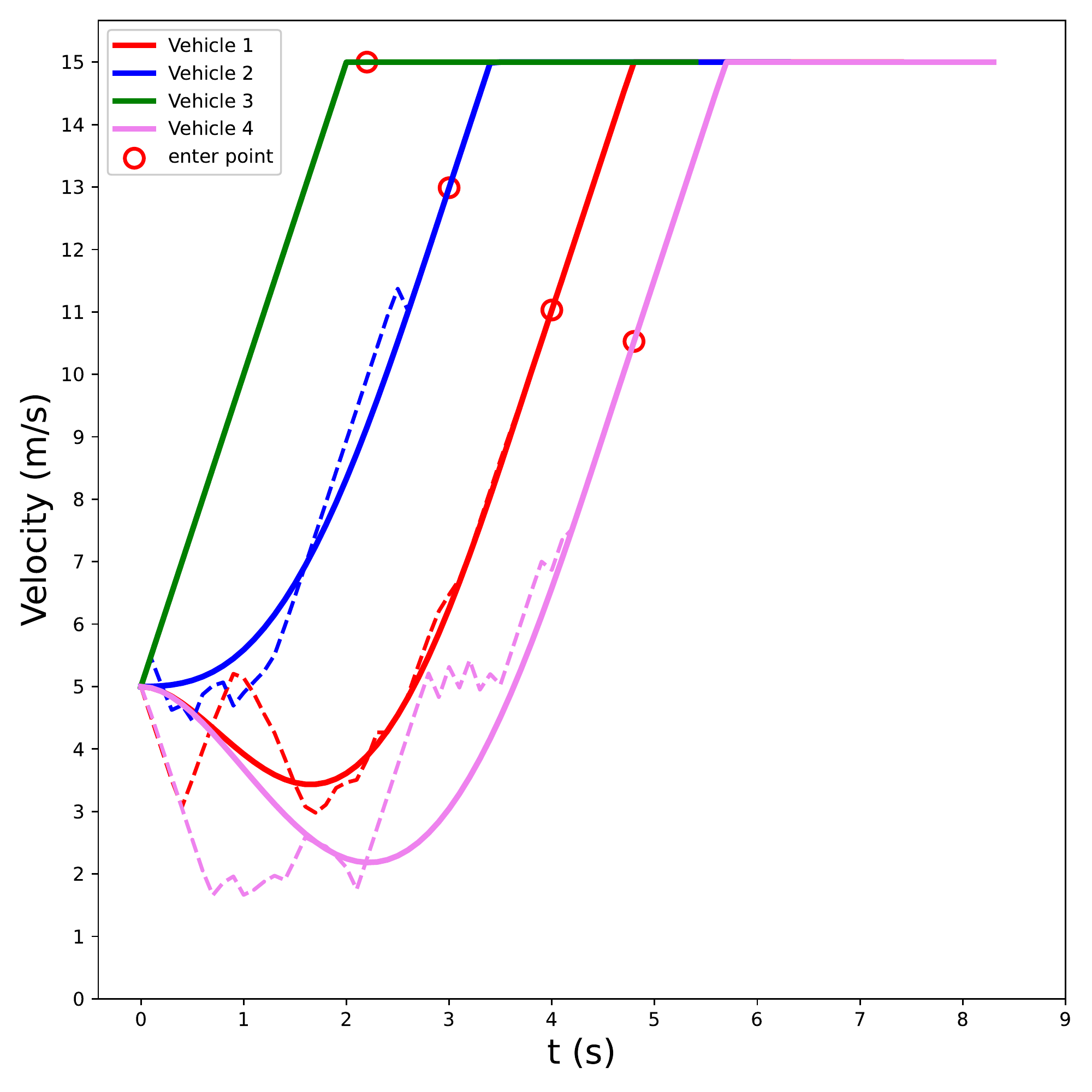}}
    \subfloat[longitudinal distance\label{fig:ST}]{%
        \includegraphics[scale=0.295]{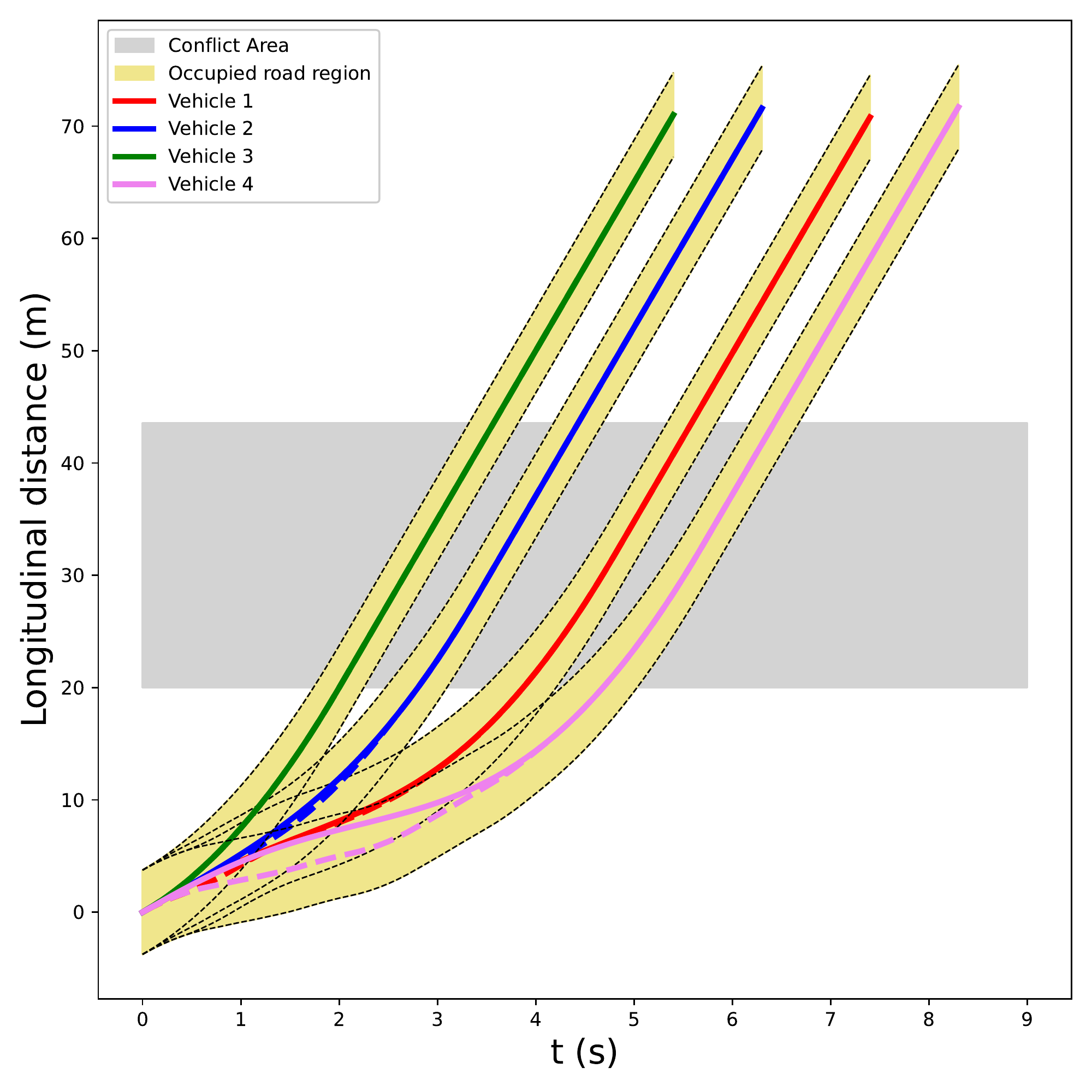}}
    \caption{The original and smoothed outputs of the TD3-based coordinator ($R_\text{l} = 15\text{m}$).}
    \label{fig:acc_vel_ST}
\end{figure*}
\subsubsection{Convergence Performance} Fig. \ref{fig:convergence_performance} demonstrates the convergence performance of the proposed TD3-based coordination strategy for single-lane scenario. As we can see from the figure, the total rewards per episode fluctuates sharply in the first 1000 episodes. As illustrated in \textbf{Algorithm \ref{algorithm_1}}, the parameters of the neural networks are randomly initialized and the agent executes random actions for better exploration in the first few episodes. Consequently, a majority of episodes terminate on a collision, resulting in unstable and undesirable rewards. However, with the parameters gradually being optimized after about 1250 episodes, the agent is intelligent enough to make proper decisions. It is obvious that the proposed TD3-based algorithm can improve the coordination efficiency significantly with a considerable convergence rate.

\subsubsection{Trajectory Analysis}
Fig. \ref{fig:CS_acc_vel_ST} and Fig. \ref{fig:acc_vel_ST} present the outputs of the CS-based and TD3-based coordinator for a static coordination scenario, respectively. All the four left-turning CVs are initialized at $p_1 = p_2 = p_3 = p_4 = 20\text{m}$ with velocity $v_\text{low} = 5\text{m/s}$ and left-turning radius $R_\text{l} = 15\text{m}$. The output information of each vehicle is distinguished by different colors, the dotted and solid lines in Fig. \ref{fig:acc_vel_ST} indicate the original and smoothed outputs, respectively. The red circles represent the entry points of the CA. It is obvious that the original acceleration and speed profile jitters dramatically before the time entering the CA. On the contrary, the smoothed profiles could provide better passenger comfort and adapt much better to the traffic flow.
Fig. \ref{fig:acc_vel_ST}(c) shows the longitudinal distance versus coordination time (i.e., S-T figure), which implies that the CA can be occupied by two vehicles simultaneously. However, as shown in Fig. \ref{fig:CS_acc_vel_ST}(c), all CVs should enter the CA one after another under the guidance of the CS strategy. Hence, the total passing time for the TD3-based strategy could be significantly reduced compared to the CS strategy.

\begin{figure*}[htp]
    \centering
    \subfloat[longitudinal distance ($R_\text{l} = 10\text{m}$)\label{fig:ST_R10}]{%
        \includegraphics[scale=0.295]{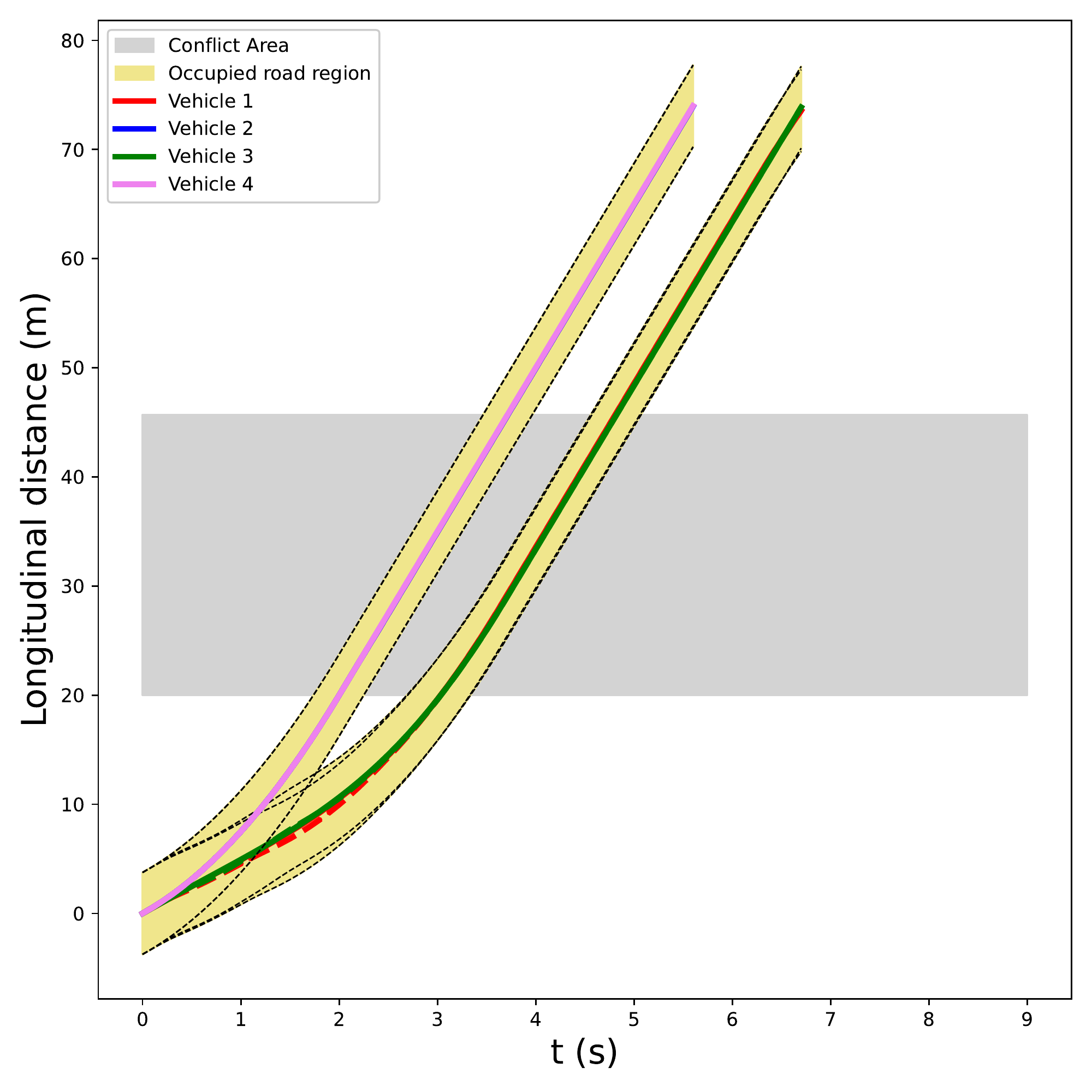}}
    \hspace{0.1cm}
    \subfloat[3D trajectories ($R_\text{l} = 15\text{m}$)\label{fig:3D_R15}]{%
        \includegraphics[scale=0.25]{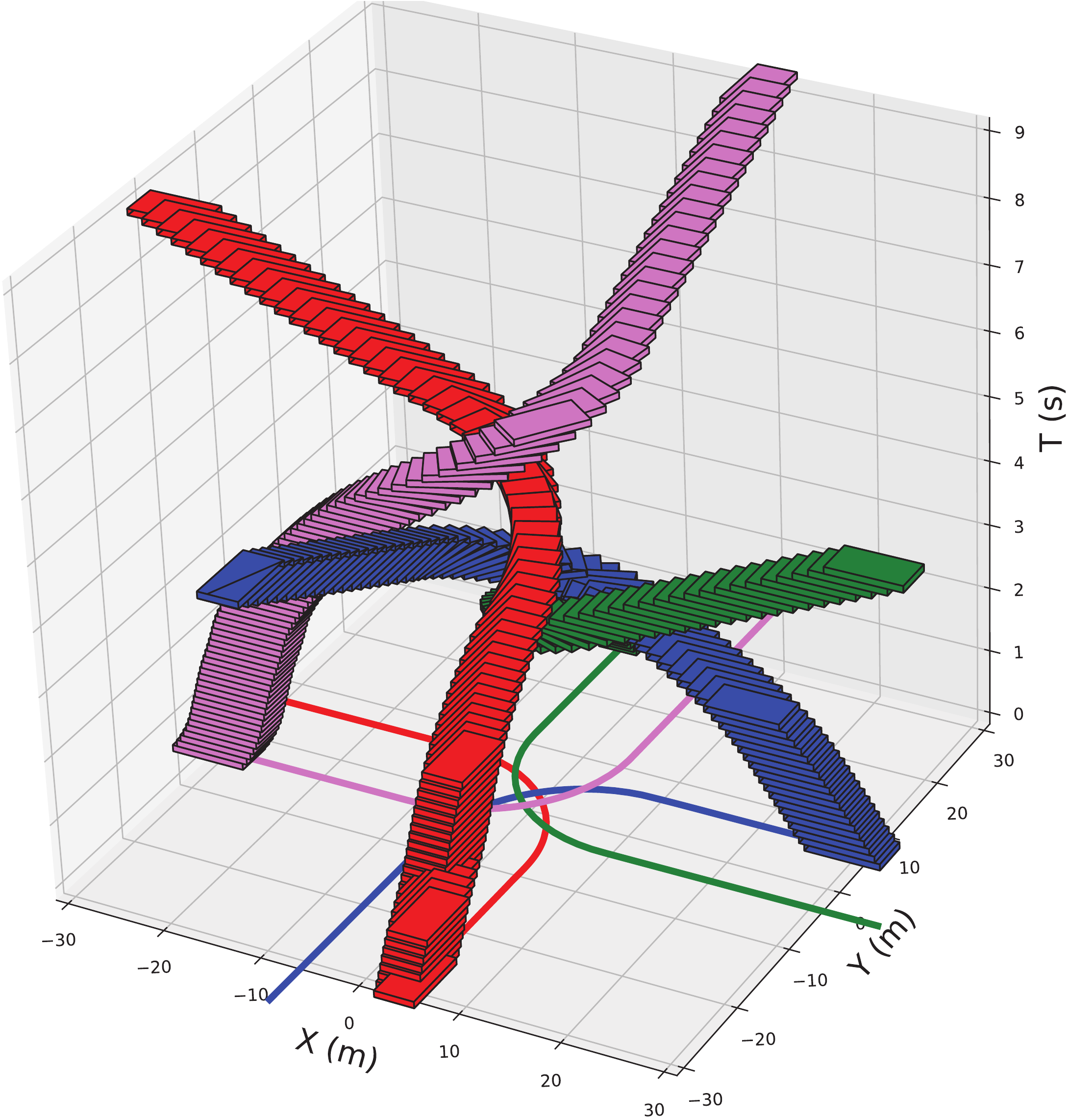}}
    \hspace{0.1cm}
    \subfloat[3D trajectories ($R_\text{l} = 10\text{m}$)\label{fig:3D_R10}]{%
        \includegraphics[scale=0.25]{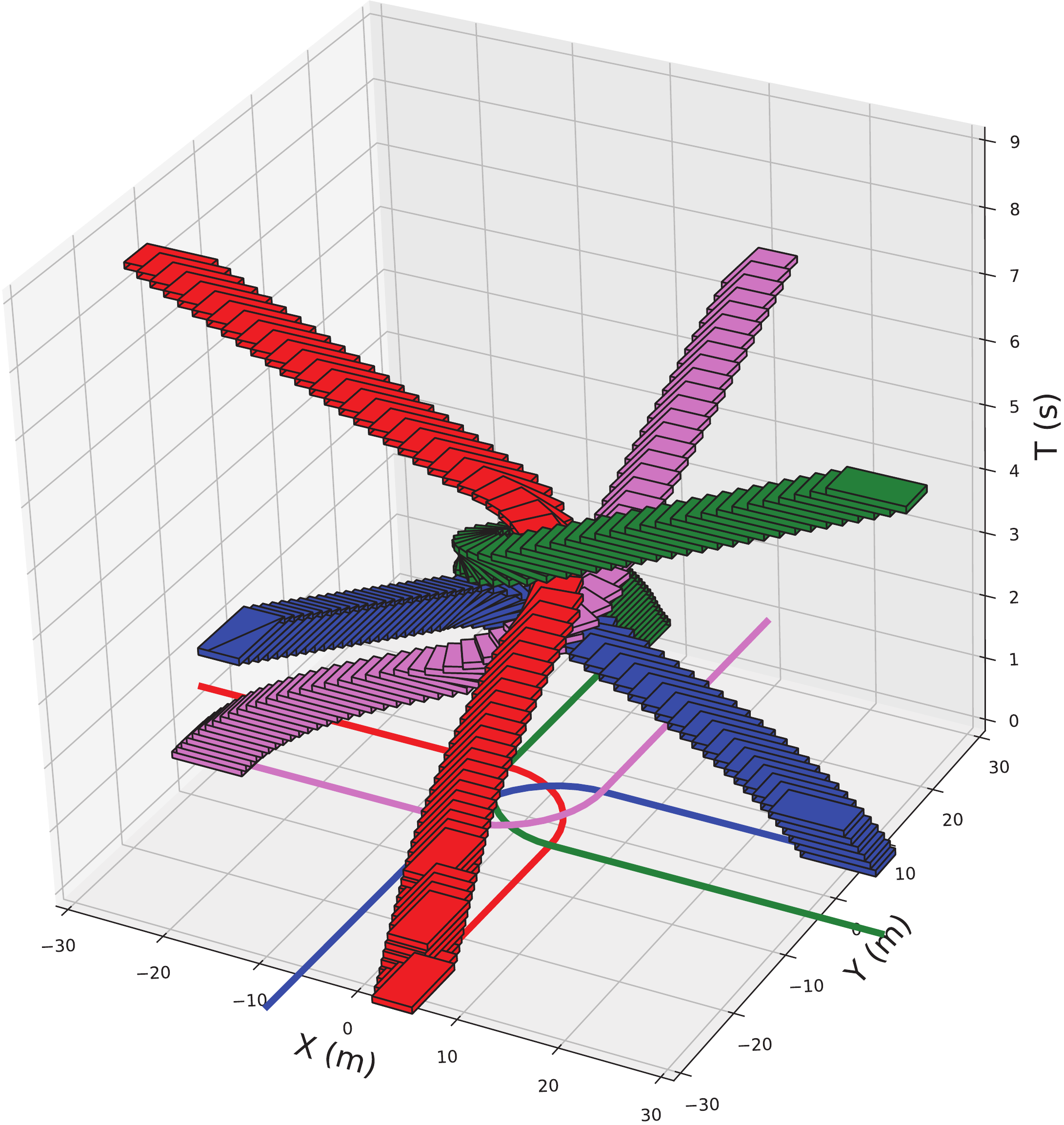}}
    \caption{Comparison of coordination efficiency for different left-turning radii.}
    \label{fig:R10_R15}
\end{figure*}

Furthermore, we recognize that the left-turning radius has a significant influence on the coordination efficiency, as shown in Fig. \ref{fig:R10_R15}. Fig. \ref{fig:R10_R15}(a) indicates that all the four CVs guided by the TD3 agent could occupy the intersection simultaneously. Fig. \ref{fig:R10_R15}(b) and Fig. \ref{fig:R10_R15}(c) show the 3D trajectories for $R_\text{l} = 15\text{m}$ and $R_\text{l} = 10\text{m}$, respectively.
From Fig. \ref{fig:R10_R15}(a) and Fig. \ref{fig:R10_R15}(c), it can be seen that the vehicles with $R_\text{l} = 10\text{m}$ in the opposite directions can cross the intersection at the same speed. In contrast, as shown in Fig. \ref{fig:acc_vel_ST}(c) and Fig. \ref{fig:R10_R15}(b), the vehicles with $R_\text{l} = 15\text{m}$ could only pass the intersection consecutively. A more detailed discussion is beyond the scope of this paper. How intersection geometry and different turning radii influence coordination efficiency will be investigated in our future work.
\begin{figure*}[htbp]
        \centering
        \begin{minipage}[t]{0.48\textwidth}
             \centering
             \includegraphics[scale=0.5]{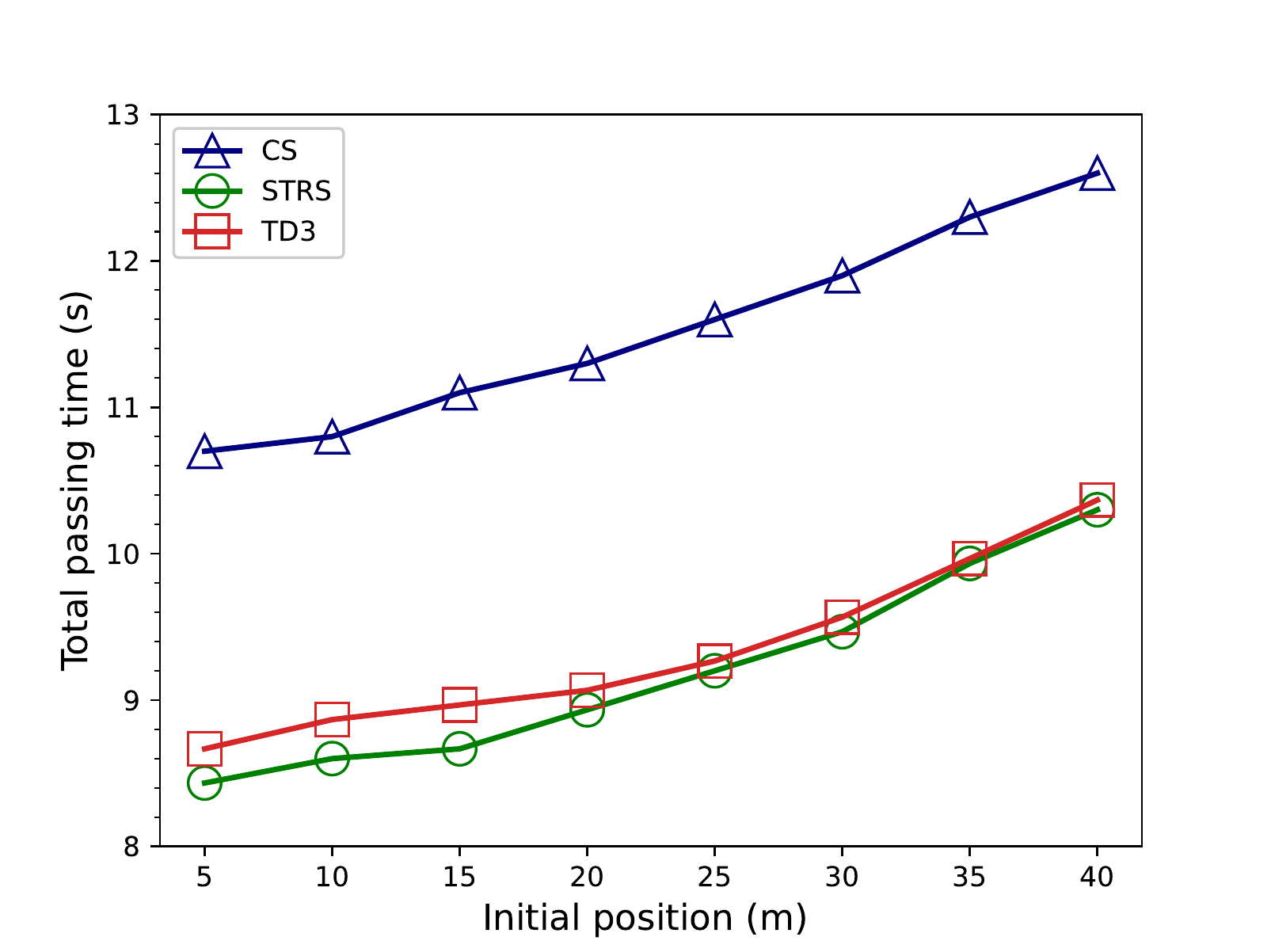}
             \caption{The total passing time of different strategies ($R_\text{l} = 15\text{m}$).}
             \label{fig:passing_time}
        \end{minipage}
        \begin{minipage}[t]{0.48\textwidth}
            \centering
            \includegraphics[scale=0.5]{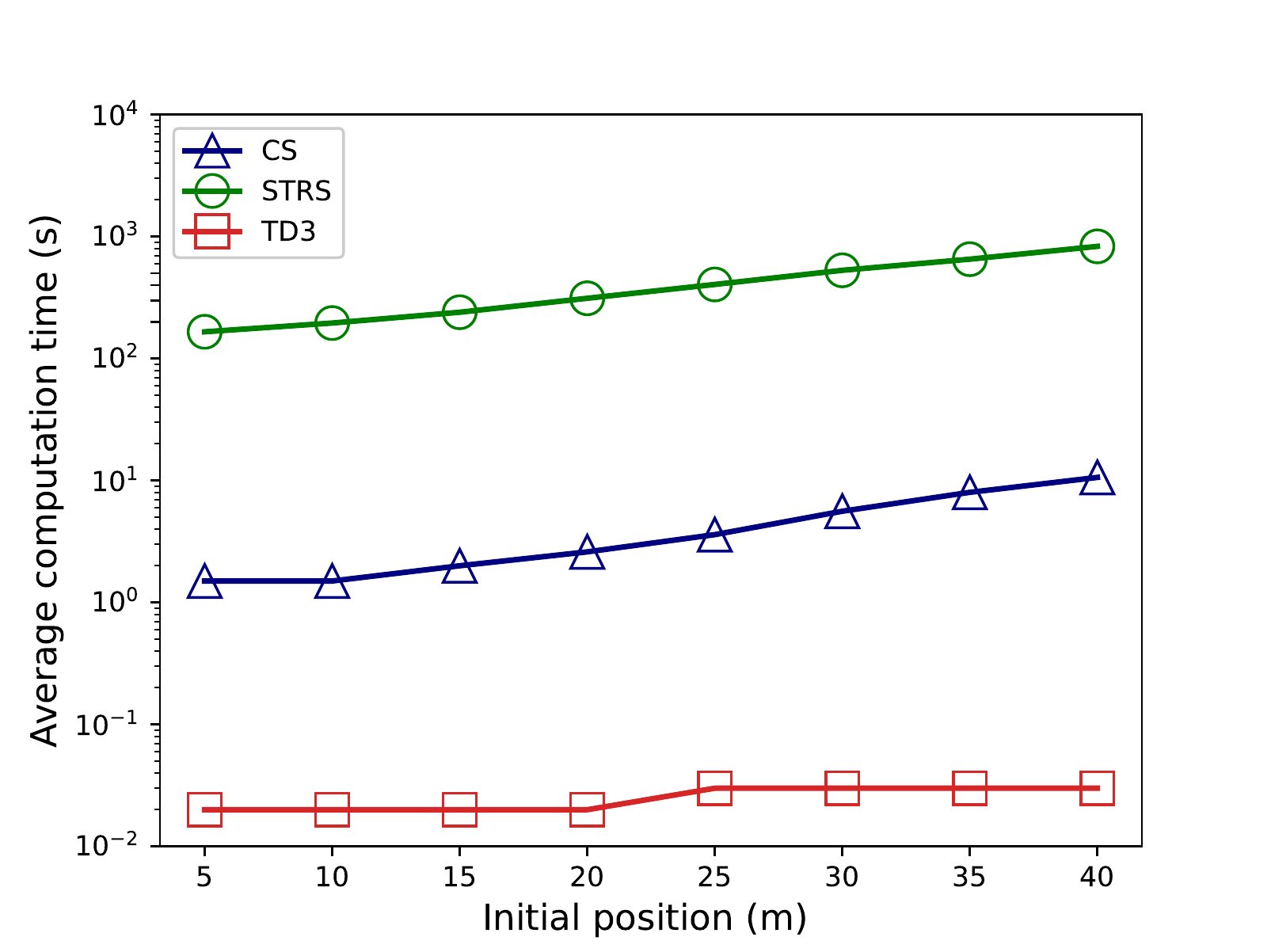}
            \caption{The average computation time of different strategies ($R_\text{l} = 15\text{m}$).}
            \label{fig:computation_time}
      \end{minipage}
\end{figure*}
\subsubsection{Total Passing Time}\label{total_passing_time}
Fig. \ref{fig:passing_time} demonstrates the total passing time of the four left-turning CVs versus the initial positions. For ease of analysis, all the CVs are initialized at $p_1 = p_2 = p_3 = p_4 = p_\text{init}$ and the left-turning radius is set as $R_\text{l} = 15\text{m}$. The total passing time of the CS strategy is much higher than that of the TD3 and STRS strategies. The STRS strategy iteratively searches the XYT resource blocks to obtain the optimal trajectories and passing order. Hence, the total passing time of the STRS strategy can be regarded as a lower bound. As can be seen from the figure, the total passage time of TD3-based strategy is almost the same as STRS strategy when $p_\text{init} > 20\text{m}$, and slightly larger when $p_\text{init} \leq 20\text{m}$.
A possible explanation for this might be that the TD3 agent tends to make conservative decisions when all CVs are close to the CA.

\subsubsection{Average Computation Time} An important observation in Fig. \ref{fig:computation_time} is that our proposed TD3-based strategy significantly outperforms other strategies in terms of average computation time.
Notably, we utilize GPUs to accelerate parallel computation in the training stage. However, in the evaluation stage, all strategies are compared on CPUs of the same platform for fairness.
As the initial positions become farther, the average computation time of the STRS and CS strategies increases due to the increase of the searching space. In particular, the STRS strategy is not a wise choice in practical implementations because the computation time is in the order of minutes. The CS strategy has a computation time of about 1-10 seconds due to its low algorithm complexity, however, at the expense of coordination efficiency.
Remarkably, our TD3-based strategy shows obvious advantages in terms of real-time processing performance as the average computation time is about 2-3ms.

\subsubsection{Traffic Throughput} 
\begin{figure*}[htbp]
        \centering
        \begin{minipage}[t]{0.48\textwidth}
             \centering
             \includegraphics[scale=0.5]{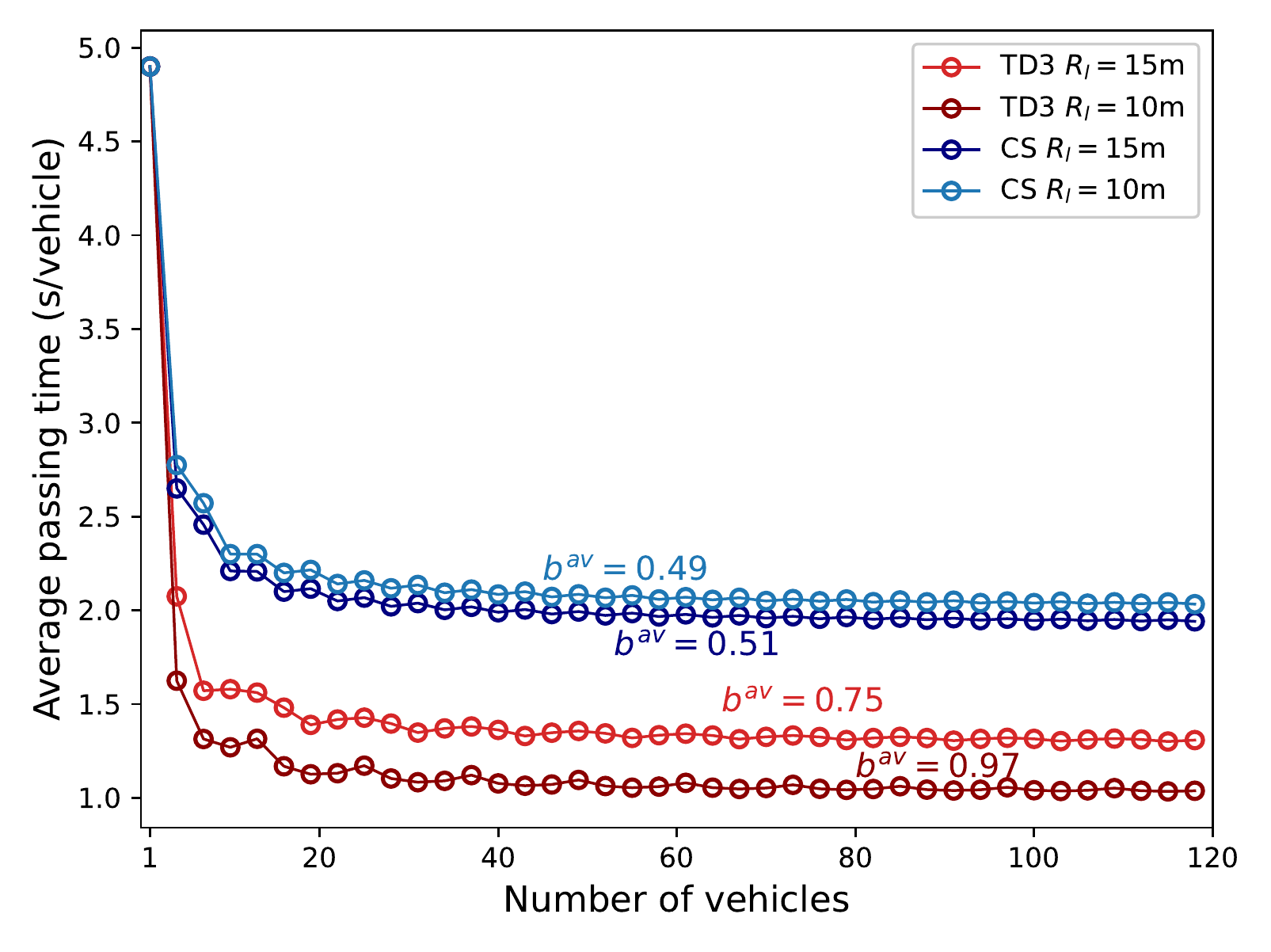}
             \caption{Traffic throughput of different strategies in single-lane scenario.}
             \label{fig:traffic_throughput}
        \end{minipage}
        \begin{minipage}[t]{0.48\textwidth}
            \centering
            \includegraphics[scale=0.5]{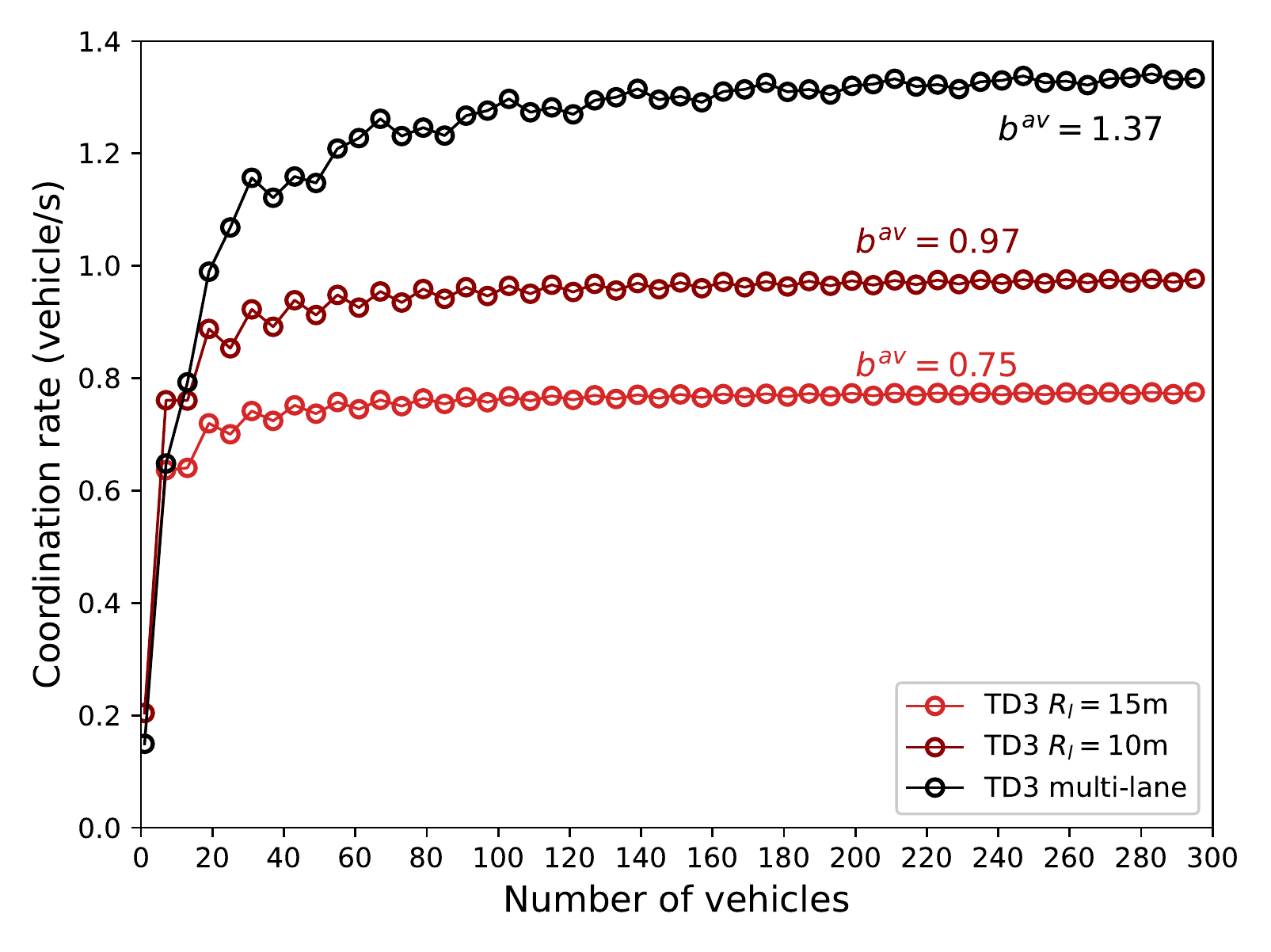}
            \caption{Traffic throughput of single-lane and multi-lane scenarios.}
            \label{fig:traffic_throughput_multi_lane}
      \end{minipage}
\end{figure*}

In Fig. \ref{fig:traffic_throughput}, we compare the traffic throughput of the CS and TD3 strategies and also present how the left-turning radius influences the traffic throughput. From the figure, the average passing time decreases with the number of waiting vehicles and turns out to be a stable value $1 / b^\text{av}$, i.e., the inverse of the coordination rate, indicating the lower the value the higher the traffic throughput. It is observed that the coordination rate of our proposed TD3-based strategy is much better than CS strategy. For $R_\text{l} = 15\text{m}$, the TD3-based strategy with 0.75 vehicle/s in mean is better than CS strategy with 0.51 vehicle/s, which means 47\% advantage in traffic throughput. Moreover, the coordination rates of CS and TD3 strategies for $R_\text{l} = 10\text{m}$ are 0.49 vehicle/s and 0.97 vehicle/s, respectively. Compared with $R_\text{l} = 15$m, there is a 4\% reduction and a 29.3\% increase in traffic throughput for the CS and TD3 strategies, respectively. This result may be explained by the fact that a small left-turning radius means more space-time resources for the TD3-based strategy but longer 2D paths for the CS strategy.

\subsection{Multi-lane Scenario}
For the CS and STRS strategies, the computational complexity grows exponentially with the number of lanes and vehicles, which presents challenges to multi-lane intersection coordination systems. In contrast, our TD3-based strategy can be easily extended to multi-lane intersections by simply adapting the input and output dimensions of the neural networks.
Fig. \ref{fig:traffic_throughput_multi_lane} compares the traffic throughput of single-lane and multi-lane scenarios. In the considered multi-lane scenario, each road has three lanes and the size of the CA is set as 48m$\times$48m. The coordination rate for the considered multi-lane intersection is 1.37 vechile/s, which means a 82.7\% and 41.2\% improvement compared to single-lane intersection with $R_\text{l} = 15$m and $R_\text{l} = 10$m. As can be seen from the figure, the traffic throughput of the three-lane intersection is less than twice of the single-lane one. This is because the CA's size of the three-lane intersection is 5.76 times larger than that of the single-lane intersection, resulting in the increase in coordination time.

\subsection{Adaptability of the proposed method}\label{batch_size}
By utilizing the offline learning methodology and real-time online processing capability of the DRL framework, our TD3-based strategy allows a batch of CVs to share the intersection area simultaneously, which significantly improves the capacity of road intersections. Although near-optimal static coordination performance can be achieved, the optimal coordination strategy for continuous traffic flow is still worth investigating. As demonstrated in \textbf{Algorithm \ref{algorithm_2}}, the WVs need to wait for the current batch of CVs to leave the CA before being coordinated. Such a strategy could avoid potential collisions between adjacent batches of CVs, but it sacrifices coordination performances.
\begin{figure}[h]
  \centering
  \includegraphics[scale=0.48]{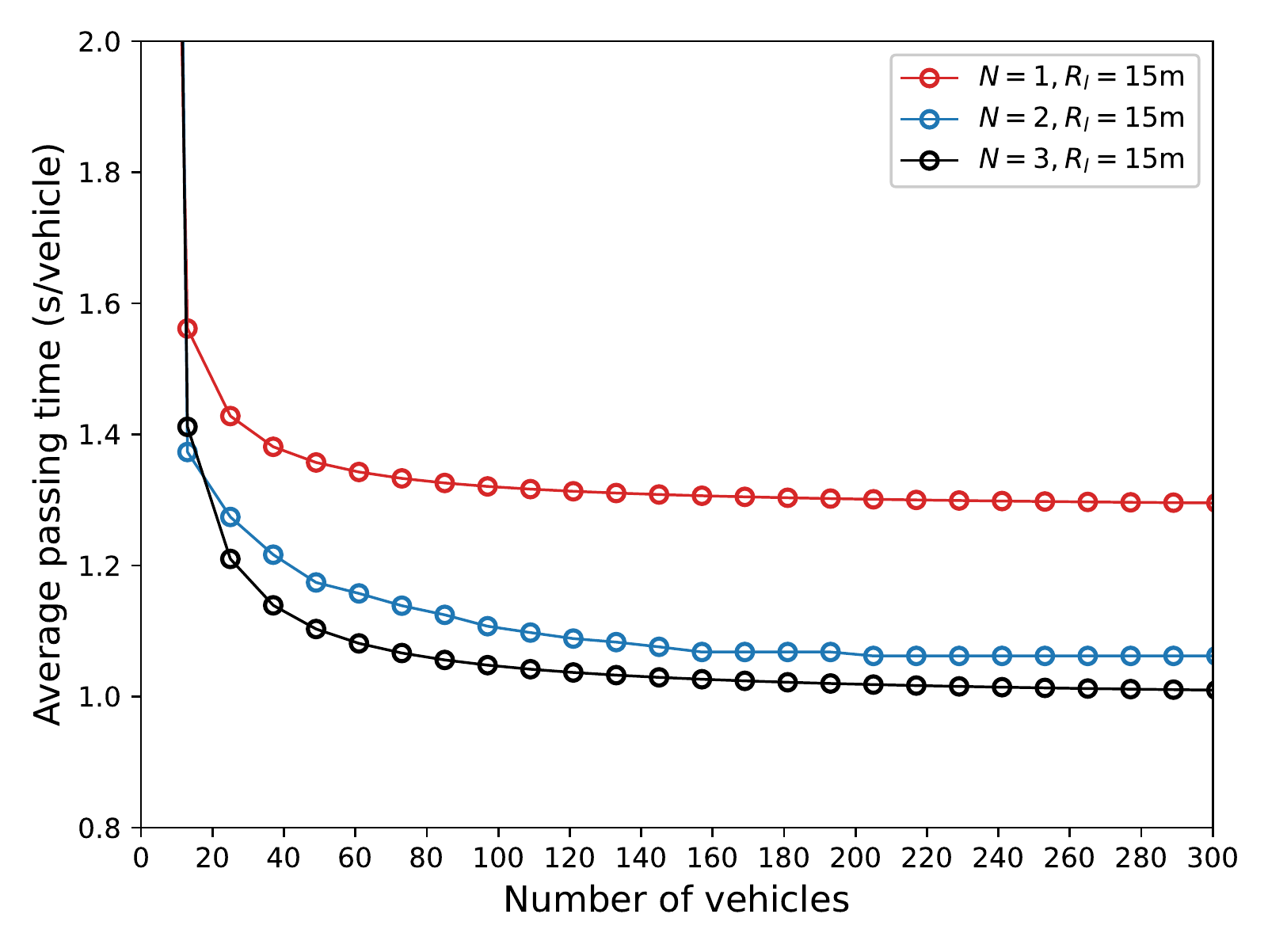}
  \caption{Coordination performance of different batch sizes.}
  \label{fig:multi_batch}
\end{figure}

One option to address this issue is to increase the batch size $N$. Fig. \ref{fig:multi_batch} evaluates the influence of different batch sizes on the average passing time in the single-lane scenario. It is obvious that a larger batch size could greatly improve coordination performance. In practice, however, we recognize that the training time grows significantly as $N$ increases (0.5h for N = 1, 1.2h for N = 2, and 3h for N = 3). Two main reasons may account for this issue: the curse of dimensionality and rear-end collision avoidance. Hence, there exists a trade-off between online coordination performance and offline training costs.
In addition, $N$ should be chosen according to the traffic loads since low-traffic intersections have a small number of WVs in the CCZ.

\begin{figure*}[htbp]
  \centering
  \includegraphics[scale=0.5]{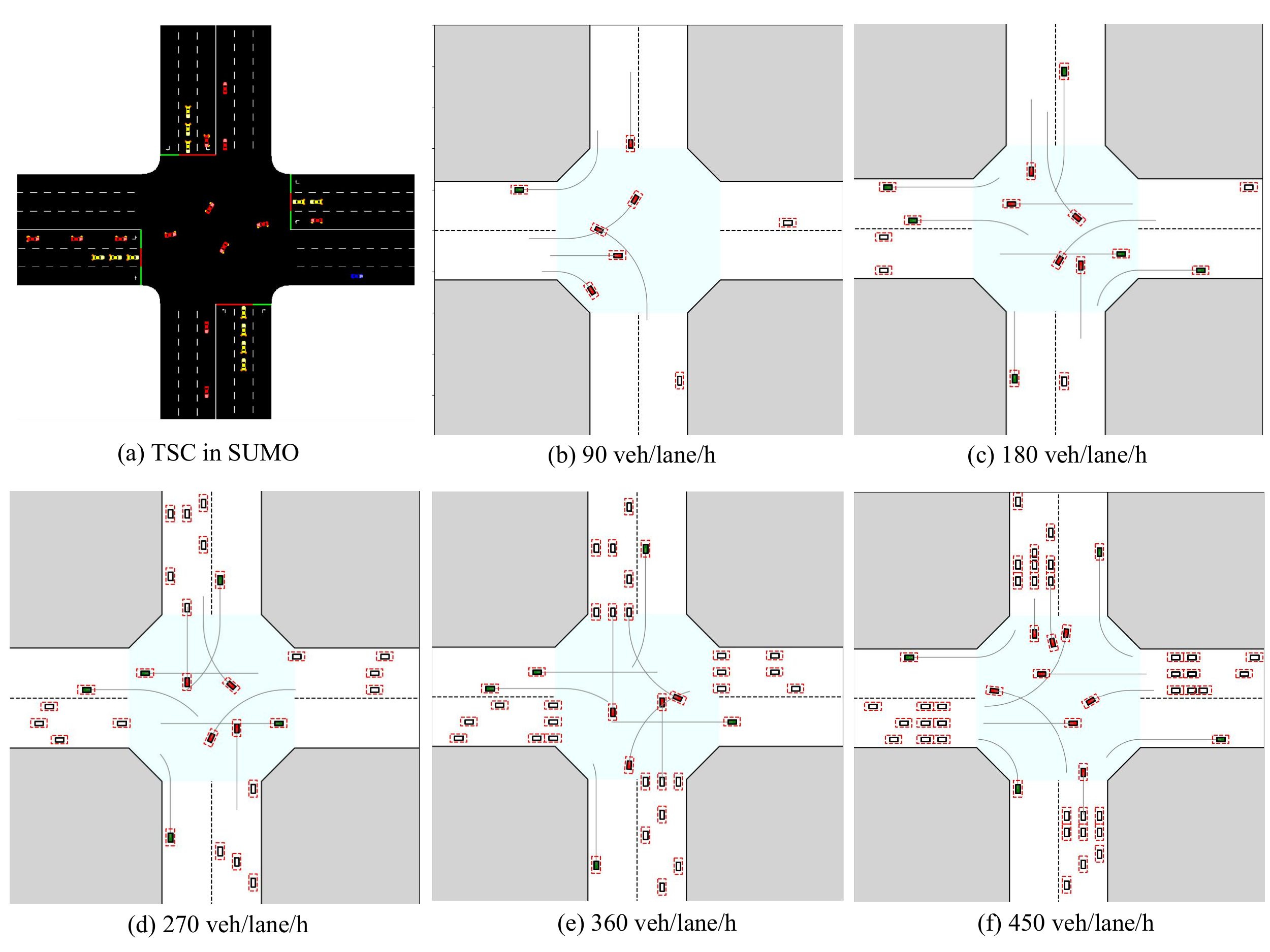}
  \caption{Simulation platforms: (a) Traditional TSC system established in SUMO. (b)-(f) Captures of cooperative coordination animation under different traffic loads with batch size $N = 1$ implemented in Python.}
  \label{fig:traffic_load}
\end{figure*}

\begin{figure}[htbp]
  \centering
  \includegraphics[scale=0.5]{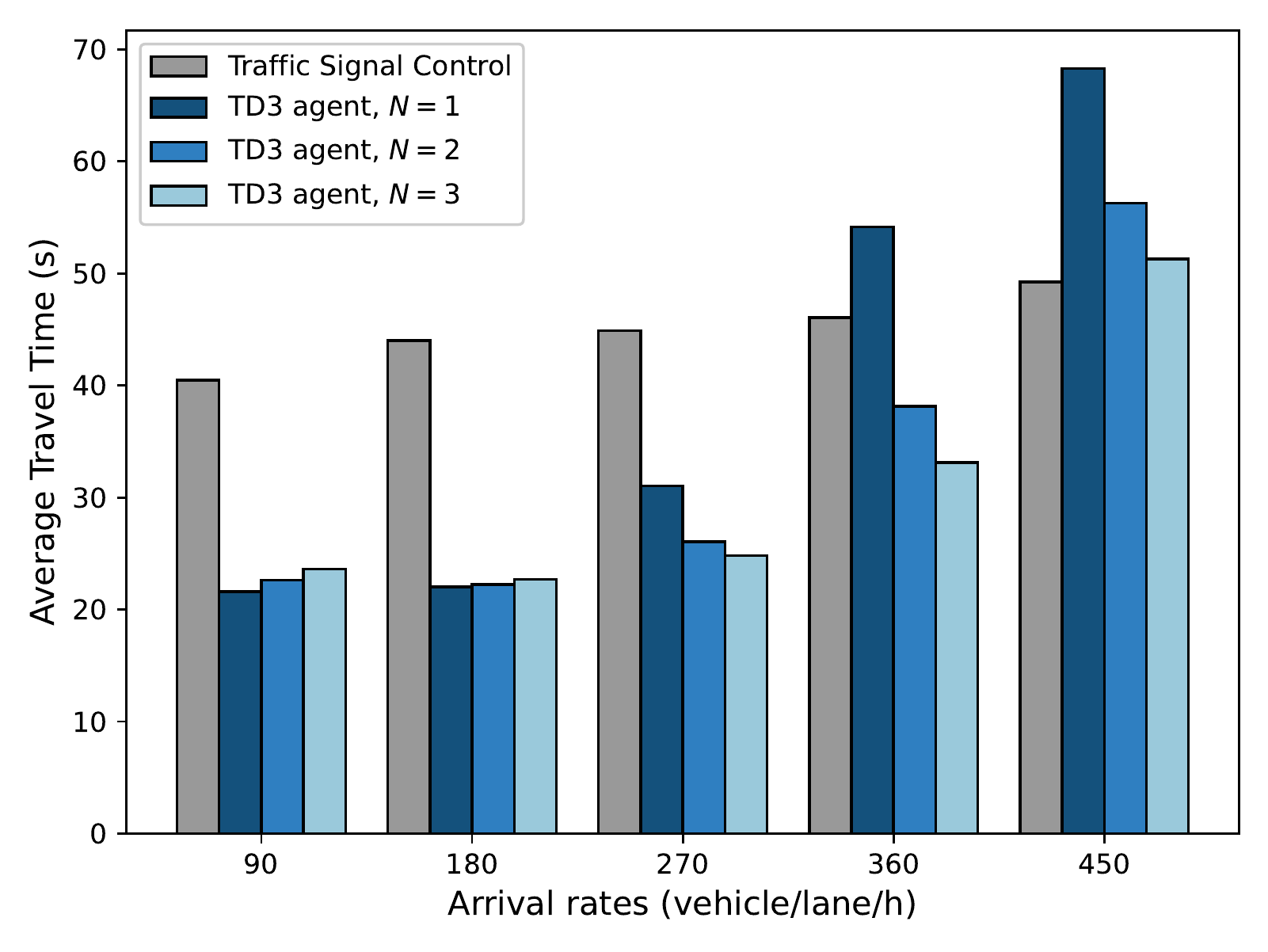}
  \caption{Comparison of average travel time under TSC and TD3 strategies.}
  \label{fig:travel_time}
\end{figure}

To evaluate the adaptability of the proposed method, we construct a multi-lane signalized intersection using the Simulation of Urban MObility (SUMO) platform. Then, we vary the arrival rates from 90 to 450 vehicles/lane/h to compare the coordination performance of the TSC system and our centralized strategy under different traffic loads, as shown in Fig.\ref{fig:traffic_load}. For each arrival rate, we simulate a 1-hour traffic process. As can be seen from Fig.\ref{fig:travel_time}, in light and medium traffic loads (90 to 270 vehicle/lane/h), our TD3 agent significantly outperforms the TSC system in terms of average travel time. It is also indicated that a larger batch size $N$ has little effect on the coordination performance since a few vehicles are within the CCZ at the same time. For the TSC system, predefined light durations could cause unnecessary waitings for the green phase, thus resulting in undesirable travel times. However, in high traffic (360 to 450 vehicles/lane/h), our strategy with $N=1$ shows unsatisfactory performance. This is because a large number of vehicles are waiting in the CCZ and the introduction of batch optimization results in more queuing time. Although a larger $N$ can greatly improve coordination performance in these cases, the traffic efficiency of our cooperative strategy under high loads is not necessarily better than the vehicle fleet under signal control. Moreover, as shown in Fig. \ref{fig:traffic_throughput_multi_lane}, the coordination rate in the multi-lane scenario is about 1.37 vehicles/s ($\approx$ 410 vehicles/lane/h), which means our TD3 agent can stabilize the system when the arrival rate is less than 410 vehicles/lane/h.

\subsection{Laboratorial Experiments}
\begin{figure}[htp]
    \centering
    \includegraphics[scale=0.28]{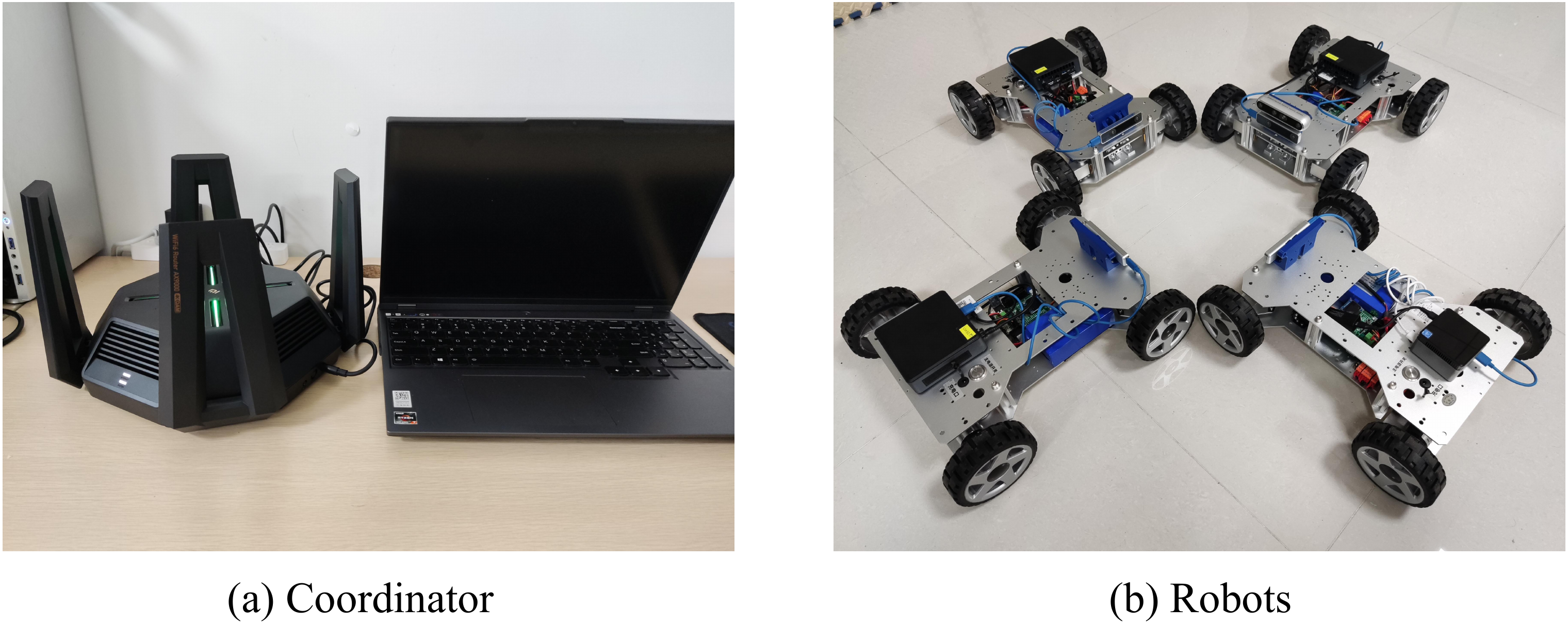}
    \caption{Devices used in the experiments.}
    \label{fig:facilities}
\end{figure}

Finally, we conduct experiments with four autonomous mobile robots based on the robot operating system (ROS)\cite{ROS} to verify the feasibility and practicability of the proposed coordination strategy. As shown in Fig. \ref{fig:facilities}, we use a laptop as the centralized coordinator and broadcast control sequences to the robots via Wi-Fi links. Each robot is equipped with a motion control unit for trajectory tracking and an Intel RealSense T265 tracking camera for self-localization. The experiment parameters are set as $v_\text{max} =0.3 \text{m/s}$, $a_\text{max} = 0.1 \text{m}/\text{s}^2$, $R_\text{l} = 1.8 \text{m}$, $\Delta T = 0.5 \text{s}$, $L_\text{car} = 0.45 \text{m}$, $W_\text{car} = 0.38 \text{m}$, $d_\text{lon} = 0.5 \text{m}$, $d_\text{lat} = 0.4 \text{m}$.

\begin{figure}[htp]
    \centering
    \includegraphics[scale=0.4]{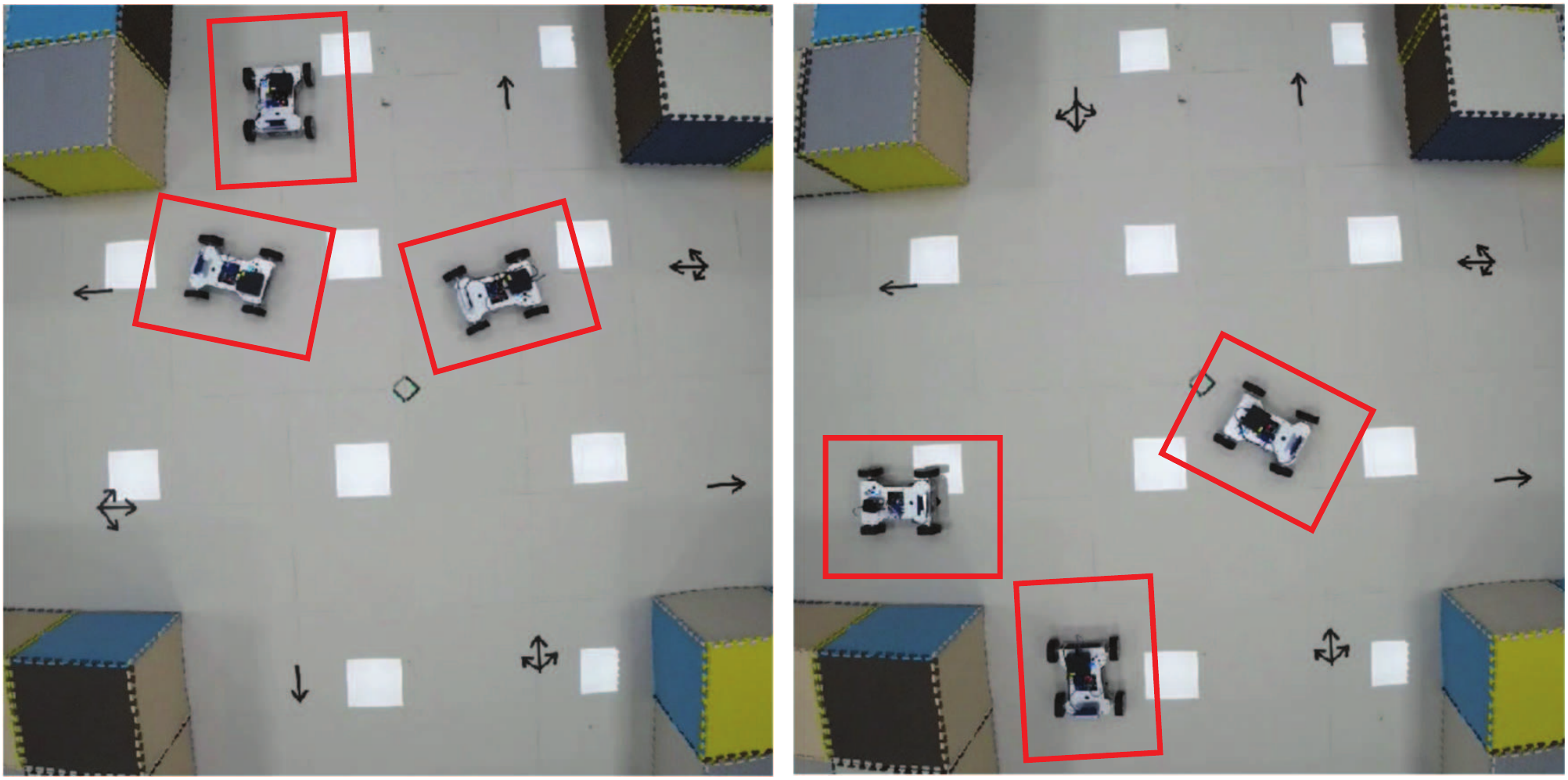}
    \caption{Laboratorial experiments based on ROS robots.}
    \label{fig:video}
\end{figure}

\begin{figure}[h]
    \centering
    \includegraphics[scale=0.28]{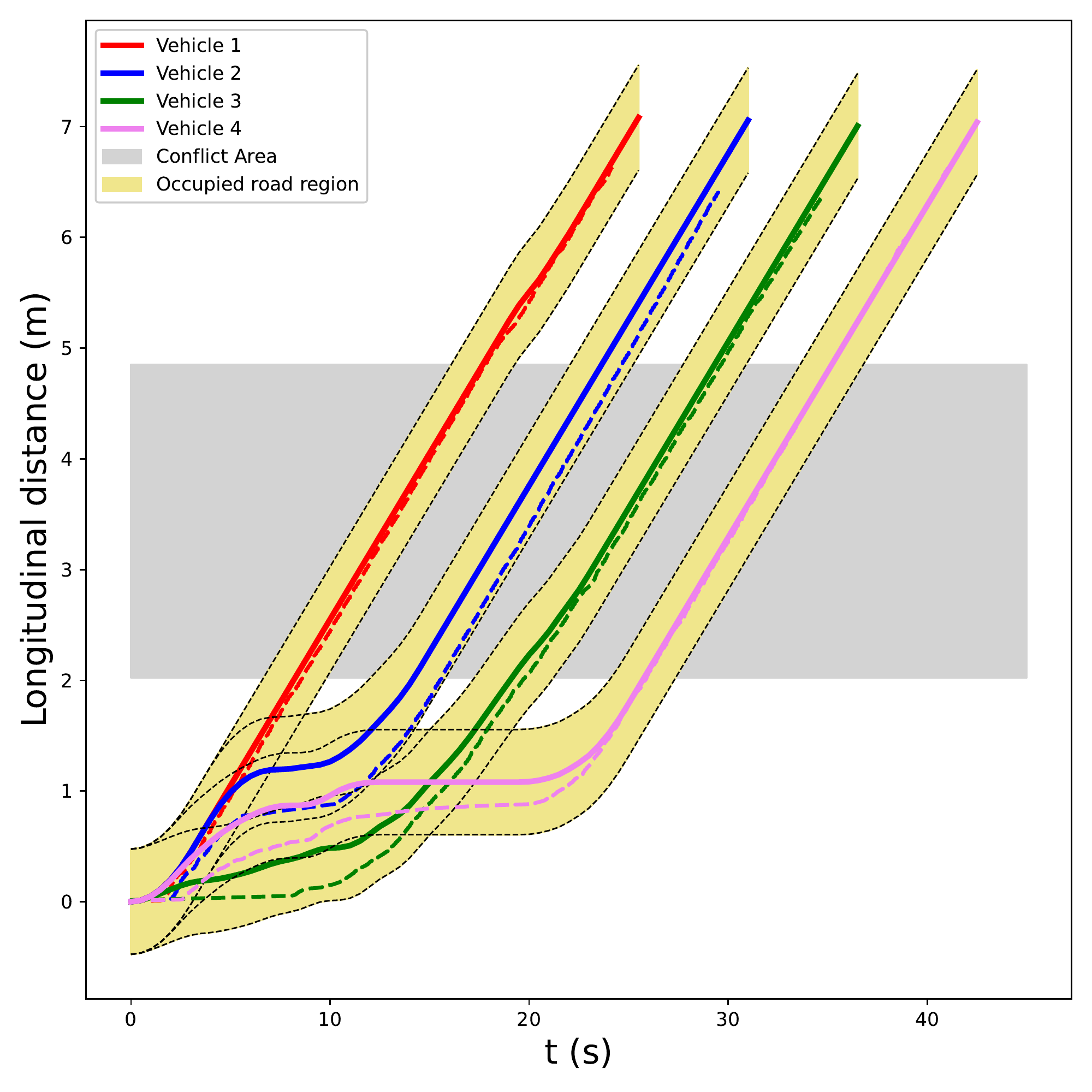}
    \caption{The deviation of planned and actual trajectories.}
    \label{fig:actual}
\end{figure}

Fig. \ref{fig:video} shows an image of the experimental coordination scenario. Our strategy enables the four left-turning robots to occupy the intersection area simultaneously, pushing the intersection coordination capacity to its limit. Fig. \ref{fig:actual} further analyzes the deviation of the planned trajectory (solid line) and actual trajectory (dashed line).
The figure shows that both the planned and actual trajectories are within the occupied road region, illustrating that the safety redundancy is well set to circumvent safety hazards caused by system imperfections. Videos of our simulations and experiments can be found online at https://www.youtube.com/watch?v=FEwa-YuFFAk.

\section{Conclusion}\label{conclusion}
In this paper, we have investigated the cooperative vehicle coordination problem at unsignalized road intersections, where the key issues to be addressed were traffic throughput, driving safety, and computational complexity. A unified optimization problem has been formulated to maximize traffic throughput while ensuring driving safety. As the original optimization problem is computationally intractable in real-time, we have exploited the DRL to transform and solve it and devised the TD3-based strategy.
Extensive experiments have shown that our strategy can significantly improve traffic throughput and have millisecond computational latency, which makes it promising to be implemented in practice.

Future research directions involve an extension of the presented framework to include a re-planning operation for accommodating system imperfections and alleviating accumulated trajectory error. Further, we will take turning radius into consideration and provide reference path alternatives to improve traffic throughput. Moreover, optimal dynamic coordination strategies in continuous traffic flow are expected to be further explored.


\bibliographystyle{IEEEtran}
\bibliography{ref}

\begin{thebibliography}{10}
\providecommand{\url}[1]{#1}
\csname url@samestyle\endcsname
\providecommand{\newblock}{\relax}
\providecommand{\bibinfo}[2]{#2}
\providecommand{\BIBentrySTDinterwordspacing}{\spaceskip=0pt\relax}
\providecommand{\BIBentryALTinterwordstretchfactor}{4}
\providecommand{\BIBentryALTinterwordspacing}{\spaceskip=\fontdimen2\font plus
\BIBentryALTinterwordstretchfactor\fontdimen3\font minus
  \fontdimen4\font\relax}
\providecommand{\BIBforeignlanguage}[2]{{%
\expandafter\ifx\csname l@#1\endcsname\relax
\typeout{** WARNING: IEEEtran.bst: No hyphenation pattern has been}%
\typeout{** loaded for the language `#1'. Using the pattern for}%
\typeout{** the default language instead.}%
\else
\language=\csname l@#1\endcsname
\fi
#2}}
\providecommand{\BIBdecl}{\relax}
\BIBdecl

\bibitem{conference_version}
J.~Luo, T.~Zhang, R.~Hao, D.~Li, C.~Chen, Z.~Na, and Q.~Zhang, ``Cooperative
  trajectory planning at unsignalized intersections using deep reinforcement
  learning,'' in \emph{2022 IEEE/CIC International Conference on Communications
  in China (ICCC Workshops)}, 2022, pp. 227--232.

\bibitem{traffic_1}
M.~Alsabaan, W.~Alasmary, A.~Albasir, and K.~Naik, ``Vehicular networks for a
  greener environment: A survey,'' \emph{IEEE Communications Surveys \&
  Tutorials}, vol.~15, no.~3, pp. 1372--1388, 2012.

\bibitem{traffic_2}
S.~Dornbush and A.~Joshi, ``Streetsmart traffic: Discovering and disseminating
  automobile congestion using vanet's,'' in \emph{2007 IEEE 65th Vehicular
  Technology Conference-VTC2007-Spring}.\hskip 1em plus 0.5em minus 0.4em\relax
  IEEE, 2007, pp. 11--15.

\bibitem{traffic_3}
G.~Marfia and M.~Roccetti, ``Vehicular congestion detection and short-term
  forecasting: a new model with results,'' \emph{IEEE Transactions on Vehicular
  Technology}, vol.~60, no.~7, pp. 2936--2948, 2011.

\bibitem{traffic_4}
K.~Mandal, A.~Sen, A.~Chakraborty, S.~Roy, S.~Batabyal, and S.~Bandyopadhyay,
  ``Road traffic congestion monitoring and measurement using active rfid and
  gsm technology,'' in \emph{2011 14th International IEEE Conference on
  Intelligent Transportation Systems (ITSC)}.\hskip 1em plus 0.5em minus
  0.4em\relax IEEE, 2011, pp. 1375--1379.

\bibitem{traffic_5}
D.~Zhao, Y.~Dai, and Z.~Zhang, ``Computational intelligence in urban traffic
  signal control: A survey,'' \emph{IEEE Transactions on Systems, Man, and
  Cybernetics, Part C (Applications and Reviews)}, vol.~42, no.~4, pp.
  485--494, 2011.

\bibitem{congesition_1}
D.~Schrank, B.~Eisele, T.~Lomax, and J.~Bak, ``Urban mobility scorecard,''
  \emph{Texas A\&M Transportation Institute}, vol.~39, p.~5, 2015.

\bibitem{accident}
E.~Bellis and J.~Page, ``National motor vehicle crash causation survey (nmvccs)
  sas analytical users manual,'' Tech. Rep., 2008.

\bibitem{TSC_2}
J.~Gao, Y.~Shen, J.~Liu, M.~Ito, and N.~Shiratori, ``Adaptive traffic signal
  control: Deep reinforcement learning algorithm with experience replay and
  target network,'' \emph{arXiv preprint arXiv:1705.02755}, 2017.

\bibitem{TSC_0}
B.~Abdulhai, R.~Pringle, and G.~J. Karakoulas, ``Reinforcement learning for
  true adaptive traffic signal control,'' \emph{Journal of Transportation
  Engineering}, vol. 129, no.~3, pp. 278--285, 2003.

\bibitem{TSC_1}
L.~Li, Y.~Lv, and F.-Y. Wang, ``Traffic signal timing via deep reinforcement
  learning,'' \emph{IEEE/CAA Journal of Automatica Sinica}, vol.~3, no.~3, pp.
  247--254, 2016.

\bibitem{TSC_3}
H.~Wei, G.~Zheng, H.~Yao, and Z.~Li, ``Intellilight: A reinforcement learning
  approach for intelligent traffic light control,'' in \emph{Proceedings of the
  24th ACM SIGKDD International Conference on Knowledge Discovery \& Data
  Mining}, 2018, pp. 2496--2505.

\bibitem{TSC_4}
X.~Liang, X.~Du, G.~Wang, and Z.~Han, ``A deep reinforcement learning network
  for traffic light cycle control,'' \emph{IEEE Transactions on Vehicular
  Technology}, vol.~68, no.~2, pp. 1243--1253, 2019.

\bibitem{DRL_ITS}
A.~Haydari and Y.~Yilmaz, ``Deep reinforcement learning for intelligent
  transportation systems: A survey,'' \emph{IEEE Transactions on Intelligent
  Transportation Systems}, 2020.

\bibitem{comfort}
P.~Dai, K.~Liu, Q.~Zhuge, E.~H.-M. Sha, V.~C.~S. Lee, and S.~H. Son,
  ``Quality-of-experience-oriented autonomous intersection control in vehicular
  networks,'' \emph{IEEE Transactions on Intelligent Transportation Systems},
  vol.~17, no.~7, pp. 1956--1967, 2016.

\bibitem{coordination_CAV_survey}
J.~Rios-Torres and A.~A. Malikopoulos, ``A survey on the coordination of
  connected and automated vehicles at intersections and merging at highway
  on-ramps,'' \emph{IEEE Transactions on Intelligent Transportation Systems},
  vol.~18, no.~5, pp. 1066--1077, 2016.

\bibitem{chenlei_1}
L.~Chen and C.~Englund, ``Cooperative intersection management: A survey,''
  \emph{IEEE Transactions on Intelligent Transportation Systems}, vol.~17,
  no.~2, pp. 570--586, 2016.

\bibitem{caodongpu_survey}
S.~Li, K.~Shu, C.~Chen, and D.~Cao, ``Planning and decision-making for
  connected autonomous vehicles at road intersections: A review,''
  \emph{Chinese Journal of Mechanical Engineering}, vol.~34, no.~1, pp. 1--18,
  2021.

\bibitem{jiapan}
Z.~Zhang, R.~Han, and J.~Pan, ``An efficient centralized planner for multiple
  automated guided vehicles at the crossroad of polynomial curves,'' \emph{IEEE
  Robotics and Automation Letters}, vol.~7, no.~1, pp. 398--405, 2021.

\bibitem{safety_gap}
J.~Lee and B.~Park, ``Development and evaluation of a cooperative vehicle
  intersection control algorithm under the connected vehicles environment,''
  \emph{IEEE transactions on intelligent transportation systems}, vol.~13,
  no.~1, pp. 81--90, 2012.

\bibitem{efficient_algorithm}
A.~Colombo and D.~Del~Vecchio, ``Efficient algorithms for collision avoidance
  at intersections,'' in \emph{Proceedings of the 15th ACM international
  conference on Hybrid Systems: Computation and Control}, 2012, pp. 145--154.

\bibitem{hult_1}
R.~Hult, G.~R. Campos, P.~Falcone, and H.~Wymeersch, ``An approximate solution
  to the optimal coordination problem for autonomous vehicles at
  intersections,'' in \emph{2015 American Control Conference (ACC)}.\hskip 1em
  plus 0.5em minus 0.4em\relax IEEE, 2015, pp. 763--768.

\bibitem{hult_2}
R.~Hult, G.~R. Campos, E.~Steinmetz, L.~Hammarstrand, P.~Falcone, and
  H.~Wymeersch, ``Coordination of cooperative autonomous vehicles: Toward safer
  and more efficient road transportation,'' \emph{IEEE Signal Processing
  Magazine}, vol.~33, no.~6, pp. 74--84, 2016.

\bibitem{moyangan}
Y.~Mo, M.~Wang, T.~Zhang, and Q.~Zhang, ``Autonomous cooperative vehicle
  coordination at road intersections,'' \emph{Journal of Communications and
  Information Networks}, vol.~4, no.~1, pp. 78--87, 2019.

\bibitem{CS}
M.~Wang, T.~Zhang, L.~Gao, and Q.~Zhang, ``High throughput dynamic vehicle
  coordination for intersection ground traffic,'' in \emph{2018 IEEE 88th
  Vehicular Technology Conference (VTC-Fall)}, 2018, pp. 1--6.

\bibitem{liuchanghao}
C.~Liu, Y.~Zhang, T.~Zhang, X.~Wu, L.~Gao, and Q.~Zhang, ``High throughput
  vehicle coordination strategies at road intersections,'' \emph{IEEE
  Transactions on Vehicular Technology}, vol.~69, no.~12, pp. 14\,341--14\,354,
  2020.

\bibitem{moyanganwcncw}
Y.~Mo, M.~Wang, T.~Zhang, and Q.~Zhang, ``Intelligent offloading strategies for
  high throughput traffic intersection coordination,'' in \emph{2019 IEEE
  Wireless Communications and Networking Conference Workshop (WCNCW)}, 2019,
  pp. 1--6.

\bibitem{CCP}
M.~A.~S. Kamal, J.-i. Imura, T.~Hayakawa, A.~Ohata, and K.~Aihara, ``A
  vehicle-intersection coordination scheme for smooth flows of traffic without
  using traffic lights,'' \emph{IEEE Transactions on Intelligent Transportation
  Systems}, vol.~16, no.~3, pp. 1136--1147, 2014.

\bibitem{lili}
H.~Xu, Y.~Zhang, L.~Li, and W.~Li, ``Cooperative driving at unsignalized
  intersections using tree search,'' \emph{IEEE Transactions on Intelligent
  Transportation Systems}, vol.~21, no.~11, pp. 4563--4571, 2019.

\bibitem{lili_2}
H.~Xu, C.~G. Cassandras, L.~Li, and Y.~Zhang, ``Comparison of cooperative
  driving strategies for cavs at signal-free intersections,'' \emph{IEEE
  Transactions on Intelligent Transportation Systems}, 2021.

\bibitem{topological_braids}
C.~Mavrogiannis, J.~A. DeCastro, and S.~S. Srinivasa, ``Implicit multiagent
  coordination at unsignalized intersections via multimodal inference enabled
  by topological braids,'' \emph{arXiv preprint arXiv:2004.05205}, 2020.

\bibitem{STRS}
Y.~Zhang, R.~Hao, T.~Zhang, X.~Chang, Z.~Xie, and Q.~Zhang, ``A trajectory
  optimization-based intersection coordination framework for cooperative
  autonomous vehicles,'' \emph{IEEE Transactions on Intelligent Transportation
  Systems}, pp. 1--15, 2021.

\bibitem{kinimatic}
P.~Hang, C.~Huang, Z.~Hu, and C.~Lv, ``Driving conflict resolution of
  autonomous vehicles at unsignalized intersections: A differential game
  approach,'' \emph{arXiv preprint arXiv:2201.01424}, 2022.

\bibitem{occupied_road_region}
R.~Firoozi, X.~Zhang, and F.~Borrelli, ``Formation and reconfiguration of tight
  multi-lane platoons,'' \emph{Control Engineering Practice}, vol. 108, p.
  104714, 2021.

\bibitem{SAT}
C.~Ericson, \emph{Real-time collision detection}.\hskip 1em plus 0.5em minus
  0.4em\relax Crc Press, 2004.

\bibitem{CBF}
H.~Xu, W.~Xiao, C.~G. Cassandras, Y.~Zhang, and L.~Li, ``A general framework
  for decentralized safe optimal control of connected and automated vehicles in
  multi-lane signal-free intersections,'' \emph{IEEE Transactions on
  Intelligent Transportation Systems}, vol.~23, no.~10, pp. 17\,382--17\,396,
  2022.

\bibitem{covariance_steering}
K.~Okamoto, M.~Goldshtein, and P.~Tsiotras, ``Optimal covariance control for
  stochastic systems under chance constraints,'' \emph{IEEE Control Systems
  Letters}, vol.~2, no.~2, pp. 266--271, 2018.

\bibitem{lyapunov}
M.~J. Neely, ``Stochastic network optimization with application to
  communication and queueing systems,'' \emph{Synthesis Lectures on
  Communication Networks}, vol.~3, no.~1, pp. 1--211, 2010.

\bibitem{silver}
D.~Silver, ``Lectures on reinforcement learning,''
  \textsc{url:}~\url{https://www.davidsilver.uk/teaching/}, 2015.

\bibitem{DQN}
V.~Mnih, K.~Kavukcuoglu, D.~Silver, A.~A. Rusu, J.~Veness, M.~G. Bellemare,
  A.~Graves, M.~Riedmiller, A.~K. Fidjeland, G.~Ostrovski \emph{et~al.},
  ``Human-level control through deep reinforcement learning,'' \emph{nature},
  vol. 518, no. 7540, pp. 529--533, 2015.

\bibitem{drone_racing}
Y.~Song, M.~Steinweg, E.~Kaufmann, and D.~Scaramuzza, ``Autonomous drone racing
  with deep reinforcement learning,'' in \emph{2021 IEEE/RSJ International
  Conference on Intelligent Robots and Systems (IROS)}.\hskip 1em plus 0.5em
  minus 0.4em\relax IEEE, 2021, pp. 1205--1212.

\bibitem{SpinningUp2018}
J.~Achiam, ``{Spinning Up in Deep Reinforcement Learning},'' 2018.

\bibitem{sutton}
R.~S. Sutton and A.~G. Barto, \emph{Reinforcement learning: An
  introduction}.\hskip 1em plus 0.5em minus 0.4em\relax MIT press, 2018.

\bibitem{TD3}
S.~Fujimoto, H.~Hoof, and D.~Meger, ``Addressing function approximation error
  in actor-critic methods,'' in \emph{International conference on machine
  learning}.\hskip 1em plus 0.5em minus 0.4em\relax PMLR, 2018, pp. 1587--1596.

\bibitem{DDPG}
T.~P. Lillicrap, J.~J. Hunt, A.~Pritzel, N.~Heess, T.~Erez, Y.~Tassa,
  D.~Silver, and D.~Wierstra, ``Continuous control with deep reinforcement
  learning,'' \emph{arXiv preprint arXiv:1509.02971}, 2015.

\bibitem{DPG}
D.~Silver, G.~Lever, N.~Heess, T.~Degris, D.~Wierstra, and M.~Riedmiller,
  ``Deterministic policy gradient algorithms,'' in \emph{International
  conference on machine learning}.\hskip 1em plus 0.5em minus 0.4em\relax PMLR,
  2014, pp. 387--395.

\bibitem{frenet}
M.~Werling, J.~Ziegler, S.~Kammel, and S.~Thrun, ``Optimal trajectory
  generation for dynamic street scenarios in a frenet frame,'' in \emph{2010
  IEEE International Conference on Robotics and Automation}.\hskip 1em plus
  0.5em minus 0.4em\relax IEEE, 2010, pp. 987--993.

\bibitem{ROS}
M.~Quigley, K.~Conley, B.~Gerkey, J.~Faust, T.~Foote, J.~Leibs, R.~Wheeler,
  A.~Y. Ng \emph{et~al.}, ``Ros: an open-source robot operating system,'' in
  \emph{ICRA workshop on open source software}, vol.~3, no. 3.2.\hskip 1em plus
  0.5em minus 0.4em\relax Kobe, Japan, 2009, p.~5.

\end{thebibliography}

\end{document}